\documentclass[fleqn,usenatbib]{mnras}

\usepackage{newtxtext,newtxmath}
\DeclareMathOperator*{\dprime}{\prime \prime}
\usepackage{ulem}

\usepackage[T1]{fontenc}

\DeclareRobustCommand{\VAN}[3]{#2}
\let\VANthebibliography\thebibliography
\def\thebibliography{\DeclareRobustCommand{\VAN}[3]{##3}\VANthebibliography}

\usepackage{graphicx}	
\usepackage{amsmath}	
\usepackage{amssymb}	

\title[Constraining dark matter with strong lensing]{A forward-modelling method to infer the dark matter particle mass from strong gravitational lenses}

\author[Q. He et al.]{Qiuhan He,$^{1}$\thanks{E-mail: qiuhan.he@durham.ac.uk}
Andrew Robertson,$^{1}$
James Nightingale,$^{1}$
Shaun Cole,$^{1}$
Carlos S. Frenk,$^{1}$
\newauthor
Richard Massey,$^{1}$
Aristeidis Amvrosiadis,$^{1}$
Ran Li$^{2,3}$,
Xiaoyue Cao$^{3,2}$, Amy Etherington$^{1}$\\
$^{1}$Institute for Computational Cosmology, Department of Physics, Durham University, South Road, Durham DH1 3LE, UK\\
$^{2}$National Astronomical Observatories, Chinese Academy of Sciences, 20A Datun Road, Chaoyang District, Beijing 100012, China\\
$^{3}$School of Astronomy and Space Science, University of Chinese Academy of Sciences, Beijing 100049, China\\
}

\date{Accepted XXX. Received YYY; in original form ZZZ}

\pubyear{2015}

\begin{document}
\label{firstpage}
\pagerange{\pageref{firstpage}--\pageref{lastpage}}
\maketitle

\begin{abstract}
  A fundamental prediction of the cold dark matter (CDM) model of
  structure formation is the existence of a vast population of dark
  matter haloes extending to subsolar masses. By contrast, other dark matter models, such as a warm
  thermal relic (WDM), predict a cutoff in the
  mass function at a mass which, for popular models, lies approximately between $10^7$ and $10^{10}~{\rm M}_\odot$. We use mock
  observations to demonstrate the viability of a forward modelling
  approach to extract information
  about  low-mass dark haloes lying along the line-of-sight to
  galaxy-galaxy strong lenses. This can be used to constrain the mass
  of a thermal relic dark matter particle, $m_\mathrm{DM}$. With 50
  strong lenses at Hubble Space Telescope resolution and
  a maximum pixel signal-to-noise ratio of $\sim50$, the expected
  median 2$\sigma$ constraint for a CDM-like model (with a halo mass cutoff at $10^{7}~{\rm M}_\odot$) is
  $m_\mathrm{DM} > 4.10 \, \mathrm{keV}$ (50\% chance of constraining $m_{\rm DM}$ to be better than 4.10~keV). If, however, the dark matter
  is a warm particle of $m_\mathrm{DM}=2.2 \, \mathrm{keV}$, our `Approximate
  Bayesian Computation' method would result in a median estimate of $m_\mathrm{DM}$ between 1.43 and 3.21~keV. Our method can be extended to the large samples of
  strong lenses that will be observed by future telescopes, and could potentially rule out the standard CDM model of cosmogony. To aid
  future survey design, we quantify how these constraints will depend
  on data quality (spatial resolution and integration time) as well as
  on the lensing geometry (source and lens redshifts).
  
\end{abstract}




\section{Introduction}

Constraining the identity and mass of the dark matter particle is one
of the most challenging goals in physics today. Astrophysics can play
a major role in this endeavour. Cosmological N-body simulations have shown
how dark matter haloes form from initial density fluctuations in the
Universe \citep{Frenk1985}. The halo mass function (i.e. the abundance
as a function of mass) and internal density profiles are determined by
the power spectrum of initial fluctuations, which itself depends on
the nature of the dark matter particle.

In the case of cold dark matter (CDM), the halo mass function extends as a
power law to very low masses (about Earth mass for a 100~GeV particle)
\citep{Wang2020} . This is perhaps the most fundamental prediction of
the standard model of cosmology, $\Lambda$CDM. By contrast, if the
dark matter is ``warm'' (WDM), the thermal velocities of the particles
at early times damp small-mass fluctuations leading to a cutoff in
the halo mass function at a scale that depends on the inverse of the
particle mass; for particles of keV mass, this is a few times 
$10^8$~M$_\odot$. Thus, a measurement of the small-mass end of the
halo mass function would set strong constraints on the properties of
the dark matter particle, including its mass, and could, in principle,
rule out some of the main current candidates, including CDM.

However, measuring the properties of dark matter haloes is not
trivial. In sufficiently massive haloes, gas can flow to the centre
and give birth to a luminous galaxy, thus, in principle, enabling
direct number counts. The observed clustering of luminous galaxies
provided the first astrophysical constraints on the mass of dark
matter particles, ruling out light particle candidates called ``hot
dark matter'', such as light neutrinos, as the predominant form of
dark matter (\citealt{Frenk1983, White1983}; see \citealt{Frenk2012}
for a review). However, these observations are unable to constrain WDM
particles which would require measurements of haloes with masses less
than $\sim$ 10$^{10}$~M$_\odot$, that are too small to have made a
luminous galaxy \citep{Thoul1996, Efstathiou1992, Benson2002,
  Sawala2016,Benitez-LLambay2020}. Furthermore, baryonic physics can
have a significant effect on dark haloes in this mass range
(particularly for those living within a galaxy) adding further
complexity to the task. Strong gravitational lensing is an observable
that can, in principle, be used to detect dark haloes and constrain
the dark matter particle mass. Other observables include the
statistics of the Lyman-$\alpha$ forest \citep{Viel2013,
  Irsic2017,Garzilli2019}, features in stellar cluster tidal debris in
the Milky Way \citep{Bonaca2019}  and populations of Milky
  Way satellites \citep{Lovell2016,Schneider2016,Lovell2017, Newton2021, Nadler2021}.

\begin{figure*}
	\includegraphics[width=2.0\columnwidth]{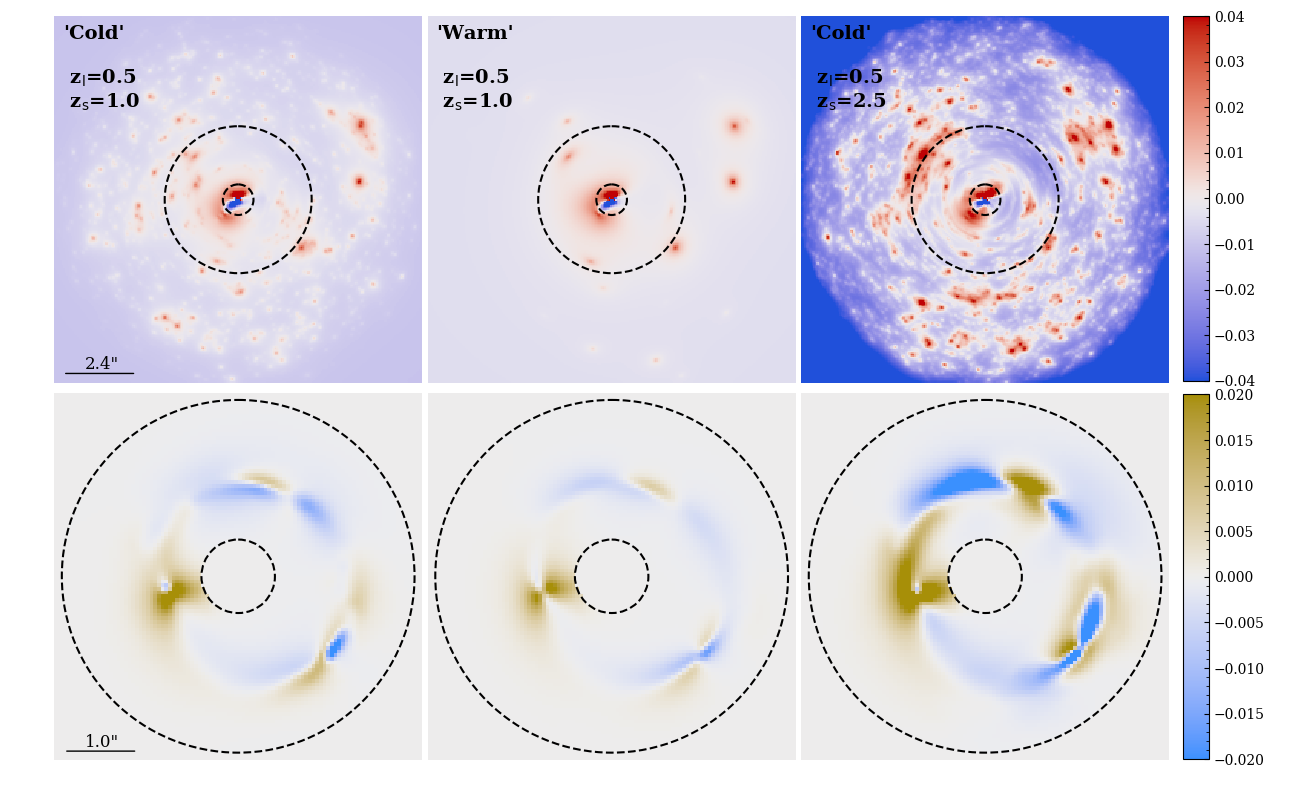}
        \caption{Upper panels: the effective convergence of low mass
          perturbers. Lower panels: the corresponding best-fit image
          residuals. The colour scale is in units of e$^{-}$
          pix$^{-1}$ s$^{-1}$. The parameters of the lens and source
          galaxies shown in the three columns are the same as those
          used in Fig.~\ref{fig:mock}, except for the source
          redshifts. The lensing systems in the left and middle
          columns have a source at $z$ = 1, while that in the right
          column has a source at $z$ = 2.5. The systems in both the
          left and right columns have a cutoff in the mass function at
          $10^7$ M$_\odot$, while the one in middle column has the
          cutoff at $10^9$ M$_\odot$. The inner and outer dashed
          circles in each panel have radii, 0.5$^{\dprime}$ and
          2.4$^{\dprime}$, respectively.}
    \label{fig:eff_cvgc}
\end{figure*}

Strong gravitational lensing is the phenomenon whereby light rays from
a distant galaxy are deflected by a foreground object, resulting in the
formation of large luminous arcs or multiple distinct galaxy
images. In principle, any object that happens to lie near the light
rays coming from the background source galaxy will leave an imprint on
the image, such that small perturbations (or lack thereof) to the
lensed source light can be used to infer the presence (or absence) of
low mass dark matter haloes. Depending on the source size and the kind
of data available, two broad types of analysis are possible. 

If the source is an extremely compact object, like a quasar, when
lensed, one will observe multiple point sources. Low mass `invisible'
dark matter haloes along the line-of-sight induce anomalies in the
flux ratios amongst multiple images, and by analysing those anomalies one can constrain the dark matter particle mass \citep{Mao1998,
  Metcalf2001, Dalal2002, Amara2006, Xu2012, Nierenberg2014, Xu2015,
  Nierenberg2017, Gilman2019, Hsueh2020, Gilman2020a}. However, because of the limited amount of information contained in the fluxes and positions of multiple images of a point-like source, there are degeneracies between the smooth model assumed for the lens and the inferred number of low-mass perturbing haloes.

If the source is extended (e.g.\ an ordinary galaxy), it appears after lensing as multiple arcs or a complete Einstein
ring. Sufficiently massive dark matter haloes along the line-of-sight
perturb local regions of the arc in a way that is measurable using
techniques such as the pixelized potential correction method
\citep{Vegetti2012, Vegetti2014, Vegetti2018,
  Ritondale2019}. By combining results for individual haloes in many
systems one can draw conclusions regarding the halo mass function
\citep[e.g.][]{Vegetti2009, Li2016}. However, this explicit subhalo modelling has
demanding requirements on image quality and resolution and so far only
a few halo detections have been reported with masses below $10^{10}$ M$_\odot$ \citep{Vegetti2010, Vegetti2012, Hezaveh2016}.

The detection of individual haloes or subhaloes requires their mass
and position to be such that they produce an observable perturbation
to the lensed image of the source. The resolution and signal-to-noise
of the data thus define the `sensitivity' to halo
detection. Haloes of mass $\sim 10^8$ M$_\odot$ are detectable with Keck AO
data \citep{Vegetti2012} while haloes as light as $10^6$ M$_\odot$ are detectable with radio
data \citep{McKean2015}. Recent studies have proposed that lower masses may be accessible
by instead extracting, statistically, the cumulative perturbation of
all intervening haloes on strong lensing arcs, even though none of
these haloes could be detected individually \citep{Brewer2016}.

Most studies have focused on determining a theoretical relation
between the power spectrum of the convergence field and that of the
image residuals after fitting a smooth `macro model' to the lens
galaxy mass distribution \citep{Chatterjee2018, Rivero2018,
  Cyr-Racine2019}. There is one example using real data, in which 
\citet{Bayer2018} analysed image residuals of the system SDSS
J0252+0039, obtaining a detection of power much higher than the
prediction of the cold dark matter (CDM) model. There remain a number
of questions as to what assumptions are appropriate for this
technique. For example, to simplify the theoretical calculations, a
random Gaussian field is often assumed for the convergence,
potentially omitting multiplane lensing effects \citep{Schneider2014c,
  McCully2014}. Indeed, recent studies have shown that the expected signal in a $\Lambda$CDM universe is
dominated by line-of-sight haloes at different redshifts to the lens
galaxy as opposed to subhaloes within the lens galaxy itself
\citep{Li2016, Li2017, Despali2018, Caganseng2020}. Here, we will
focus on how the line-of-sight low mass dark matter haloes contribute
to perturbations on lensing images.

To form a visual impression of how the line-of-sight low-mass dark
haloes perturb the images, in Fig.~\ref{fig:eff_cvgc} we show the
effective convergence, as derived from the deflection angles
\citep[upper panels;][]{Gilman2019, Caganseng2020}, and the
corresponding residuals obtained by fitting the smooth lensing
model to the image (lower panels). As may be seen from the effective
convergence map, even though all low-mass dark matter haloes are
modelled as spherical profiles, some are heavily stretched
into arc-like features due to multiplane lensing effects\footnote{ In
  strong lensing, as viewed from the image plane, not only is the distant
  light warped by foreground objects, like a luminous source galaxy
  deflected by a lens, but also the mass on distant planes is warped
  by near planes, and thus we see some arc-like low-mass dark matter
  haloes in these convergence maps even though they are all
  spherical.}, which are 
difficult to compute analytically. Comparing the left two columns, the
`warm' case in the middle and the `cold' case in the left have very
similar residuals, particularly the large patches, even though the
warm case has far fewer low-mass dark matter haloes. This suggests
that the residuals are dominated by massive dark matter haloes and to
distinguish between different dark matter models, we need to identify
the small patches. The system shown in the third column has a higher
redshift source and thus many more low-mass dark matter haloes along
the line-of-sight, which results in larger and more complicated
residuals. 

In this work, we investigate whether a forward modeling procedure
built around the approximate Bayesian computation (ABC) statistical
inference method can extract meaningful information about the
properties of low-mass dark matter haloes perturbing strong lensing
images. As opposed to determining a concise mathematical expression,
our forward modelling method relies on robust modelling of the
perturbations on lensing images induced by small dark haloes which can
directly build up a relation between the models of interest and the
observations without any further assumptions. The approach was first
applied to strong lensing by \citet{Birrer2017} who analysed the
strongly lensed quasar RXJ1131-1231. Subsequently, \citet{Gilman2019}
used this approach to study flux ratio anomalies of point-like sources
with a full realisation of dark matter haloes including line-of-sight
dark haloes.

We apply the ABC framework to a simulated sample of 50 HST-resolution strongly lensed extended sources, which is a comparable number of lenses to the high-quality strong lensing SLACS sample \citep[][but the details of our sample are different from those of SLACS]{Bolton2006}. Each simulation contains a full
cosmological realisation of small dark haloes whose redshift distribution requires analysis using multiplane ray tracing. To apply the ABC method we
simulate and refit each of these 50 lenses $20\,000$ times, producing a
total of $1$ million lens models. The scale of this analysis
necessitates an automated framework for lens modelling, for which we
use the open source software
\text{PyAutoLens}\footnote{https://github.com/Jammy2211/PyAutoLens}. Our aim
is to determine whether the cumulative distortions due to the many
dark matter haloes perturbing the light of the lensing arcs can be
extracted to determine the halo mass function and hence the dark
matter particle mass.


Our main goal is to demonstrate that this signal is present in HST imaging of
strong lenses and that it can, in principle, be extracted using modern
lens modeling techniques given realistic levels of noise. However, our
study is based on idealized systems: we make a number of simplifying
assumptions for the structure of the lens and source and neglect
effects such as an imperfect PSF model, correlated noise or inadequate
lens light subtraction.  These assumptions will need to be relaxed
before our methodology can reliably be applied to real data,
including, for example, a non-parametric source model \citep[e.g.][]
{Warren2003}, additional complexity in the lens model
\citep{Vegetti2009, Nightingale2019} and a proper treatment of the PSF
and of correlated noise.

We first introduce the forward modelling procedure and lensing models
in Sec.~\ref{sec:method}. In Sec.~\ref{sec:results} we show tests of
the accuracy of this method, the dependency on different lensing and
observational settings, and compare our method to other methods,
discussing its possible future applications and shortcomings. Finally,
we conclude in Sec.~\ref{sec:conclusion}. Throughout the paper, we
adopt the cosmological parameters given by WMAP9 \citep{Hinshaw2013}.

\begin{figure*}
	\includegraphics[width=2.0\columnwidth]{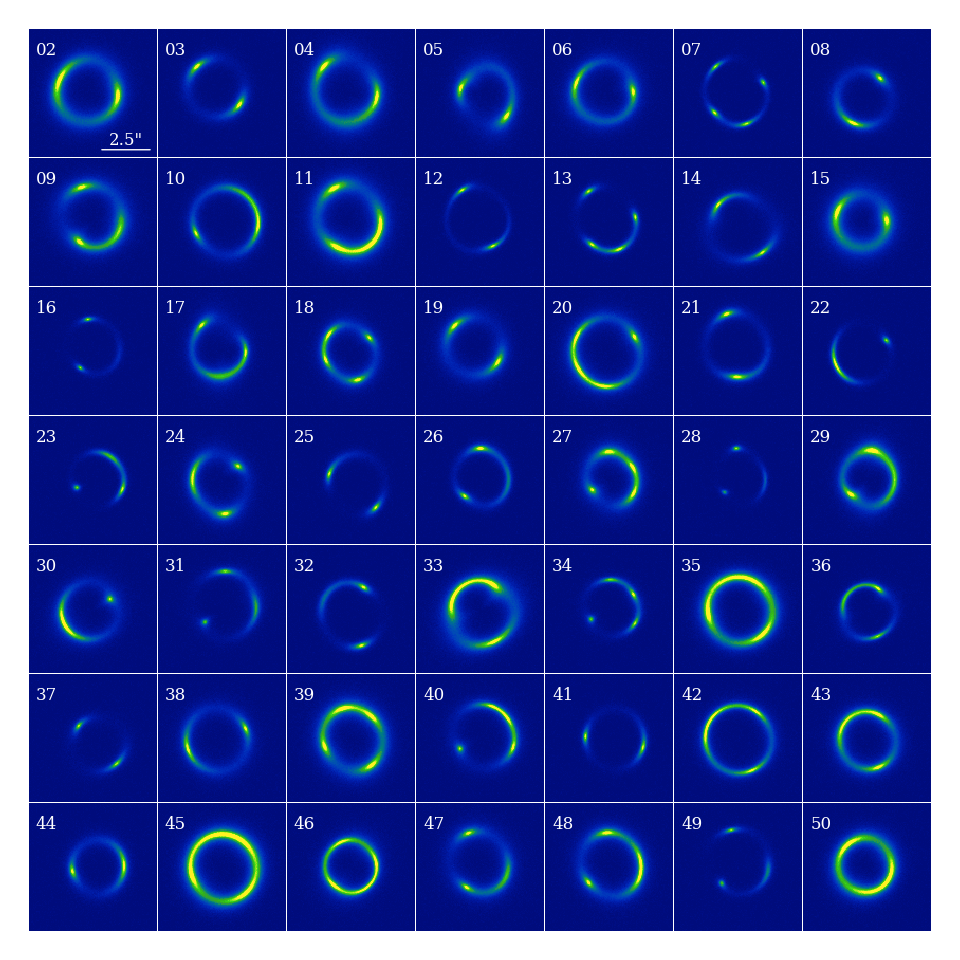}
        \caption{Fiducial mock strong lensing images. A total of 49
          mock images are shown, with the last of our set of 50 shown
          in the upper panel of Fig.~\ref{fig:mock}. The images are
          simulated with parameters randomly drawn according to
          Table~\ref{tab:drawn_rges}. }
    \label{fig:arcs} 
\end{figure*}

\section{Procedure and models}\label{sec:method}
In this section we first provide an overview of our forward modelling
procedure. We then describe the parametric models we use for the mass
distributions of the main lens, dark matter haloes and light
distributions of the source galaxies. We then describe how we fit our
simulated images with a combination of a smooth parametric lens and
source model, and how the residuals of each are used within an ABC
framework to place constraints on the mass function of dark matter
haloes. 

\subsection{The forward modelling scheme}
\begin{figure}
	\includegraphics[width=\columnwidth]{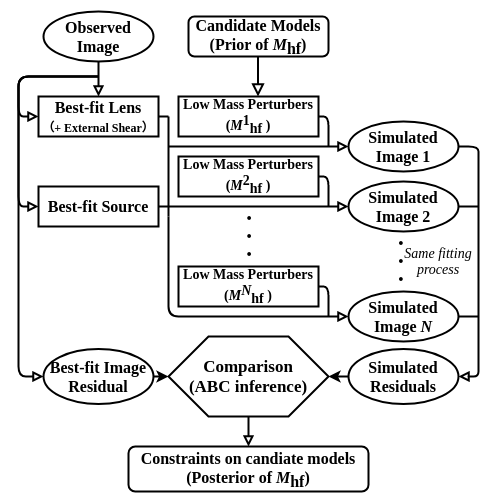}
        \caption{An overview of the forward modelling procedure for
          one lensing system. The observed image is fitted with a
          parametric lens mass model and a parametric source light
          model, producing a best-fit lens model, source model and
          image residuals. We use this best-fit model to generate $N$
          simulated images, including random realisations of low-mass
          perturbing haloes in the lens model. The number of
          perturbers as a function of mass depends on the properties
          of the dark matter as encoded in the `half-mode mass',
          $M_\mathrm{hf}$. The distribution of $M_\mathrm{hf}$ over
          the $N$ `forward models' follows our prior on
          $M_\mathrm{hf}$. Each of the forward model images is fit in
          the same way as the observed image to produce $N$ sets of
          image residuals. These forward model residuals are compared
          with the observed residuals, using ABC to constrain
          $M_\mathrm{hf}$.}
    \label{fig:fwmd}
\end{figure}

In Fig.~\ref{fig:fwmd}, we provide an overview of the forward modelling
procedure. Starting from an observed strong lensing image (which in
this paper is simulated) we begin by fitting it with parametric lens
and source models, omitting substructure from the lens model. This
procedure gives us best-fit smooth lens mass and source light models,
as well as a map of the best-fit image residuals (the observed image
minus the best-fit model image). In this work, `best-fit' refers to
the maximum likelihood model determined by means of a non-linear
search.

It is impossible to fit all the dark matter substructre in a similar way, because of its low mass and low signal-to-noise \citep{Birrer2017}. However, we can use the best-fit source and macroscopic lens model to simulate a
set of images of this lens system, each including a random
realization of dark matter substructure (we call this set of images
the `forward models'). We refit each forward model in the same manner
as we fitted the `observed' image, providing best-fit image residuals
for each forward model. We compare the forward-modelled residuals with
the observed residuals, as described in detail in Sec~\ref{sec:mocks}, and
apply ABC inference to obtain a constraint on the cutoff in the halo mass function.

\begin{table}
	\centering
	\begin{tabular}{cc} 
	    \hline
	    Parameter & Range or Value \\
	    \hline
	    \textbf{Lens} & Elliptical Isothermal \\
	    centre (x, y) [$^{\dprime}$] & (0.0, 0.0) \\
	    $R_{\rm E}$ [$^{\dprime}$] & [1.2, 1.6] \\
	    axis ratio & [0.6, 0.95] \\
	    position angle [$^\circ$] & 30 \\
	    redshift & 0.5 \\
	    \hline
	    \textbf{External Shear} & \\
	    magnitude $A$ & 0 \\
	    position angle [$^\circ$] & 0 \\
	    \hline
	    \textbf{Source} & Elliptical Core-Sersic \\
	    centre (x, y) [$^{\dprime}$] & ([$-0.3$, 0.3], [$-0.3$, 0.3]) \\
	    r$_{\rm e}$ [$^{\dprime}$] & [0.1, 0.5] \\
	    n & [1.5, 2.5] \\
	    I$^{'}$ [e$^{-}$ pix$^{-1}$ s$^{-1}$] & 7.4 \\
	    axis ratio & [0.5, 0.95] \\
	    position angle [$^\circ$] & [10, 100] \\
	    r$_{\rm c}$ [$^{\dprime}$] & 0.01 \\
	    $\alpha$ & 2.0 \\
	    $\gamma$ & 0.0 \\
	    redshift & 1.0 \\
	    \hline
	    \textbf{Dark haloes} & Truncated NFW profile \\
	    $F_{\rm CDM}$ & \citet{Sheth2001} \\
	    $c_{\rm CDM}$ & \citet{Ludlow2016} \\
	    $\sigma_{\rm log_{10}\ c_{\rm CDM}}$ & 0.15 dex \\
	    tNFW $\tau$ & $r_{100} / r_\mathrm{s}$ \\
	    M$_{\rm hf}$ [M$_\odot$] & 10$^7$ \\
	    \hline
	    \textbf{Image} & \\
	    pixel size [$^{\dprime}$] & 0.05 \\
	    PSF $\sigma$ [$^{\dprime}$] & 0.05 \\
	    t$_{\rm exp}$ [s] & 600 \\
	    background sky [e$^{-}$ pix$^{-1}$ s$^{-1}$] & 1.0 \\
	    max S/N of pixels & $\sim$ 50 \\
	    \hline
	\end{tabular}
	\caption{Fiducial model parameters. Closed brackets indicate
          the range from which values are randomly
          drawn.}\label{tab:drawn_rges}
\end{table}
\begin{table}
    \renewcommand\arraystretch{1.5}
	\centering
	\begin{tabular}{ccc} 
		\hline
		& $\{ T \}^1_{\rm fiducial}$ & $\{ F \}^1_{\rm fiducial}$ \\
		\hline
		\textbf{Lens} & \multicolumn{2}{c}{Elliptical Isothermal} \\
		centre (x, y) [$^{\dprime}$] & (0.0, 0.0) & (-0.001$_{-0.002}^{+0.002}$, -0.001$_{-0.003}^{+0.003}$) \\
		$R_{\rm E}$ [$^{\dprime}$] & 1.342 & 1.342$^{+0.002}_{-0.003}$ \\
		axis ratio & 0.618 & 0.617$^{+0.005}_{-0.005}$ \\
		position angle [$^\circ$] & 30.0 & 30.0$^{+0.4}_{-0.4}$\\
		\hline
		\textbf{External Shear} & & \\
		magnitude & 0 & 0.002$^{+0.002}_{-0.002}$ \\
		position angle [$^\circ$] & 0 & 159$^{+37}_{-41}$ \\ 
		\hline
		\textbf{Source} & \multicolumn{2}{c}{Elliptical Core-Sersic} \\
		centre (x, y) [$^{\dprime}$] & ($-0.072$, $0.259$) & ($0.071^{+0.001}_{-0.001}$, $0.253^{+0.001}_{-0.001}$) \\
		$r_{\rm e}$ [$^{\dprime}$] & 0.244 & $0.240^{+0.005}_{-0.004}$ \\
		$n$ & 1.85 & $1.81^{+0.05}_{-0.05}$ \\
		$I^{'}$ [e$^{-}$ pix$^{-1}$ s$^{-1}$] & 7.4 & $7.3^{+0.2}_{-0.2}$ \\
		axis ratio & 0.85 & $0.85^{+0.01}_{-0.01}$ \\
		position angle [$^\circ$] & 41 & $40^{+2}_{-2}$ \\
		\hline
	\end{tabular}
	\caption{The true input and best-fit parameters for the first
          lensing system. The errors listed in this table are
          3-$\sigma$ limits. Notice that we fit a system with additional low-mass perturbers, and thus the inconsistency of the best-fit parameters and the true values is expected. Parameters fixed during fitting are not
          shown here.}\label{tab:mock_paras} 
\end{table}
\begin{figure}
	\includegraphics[width=\columnwidth]{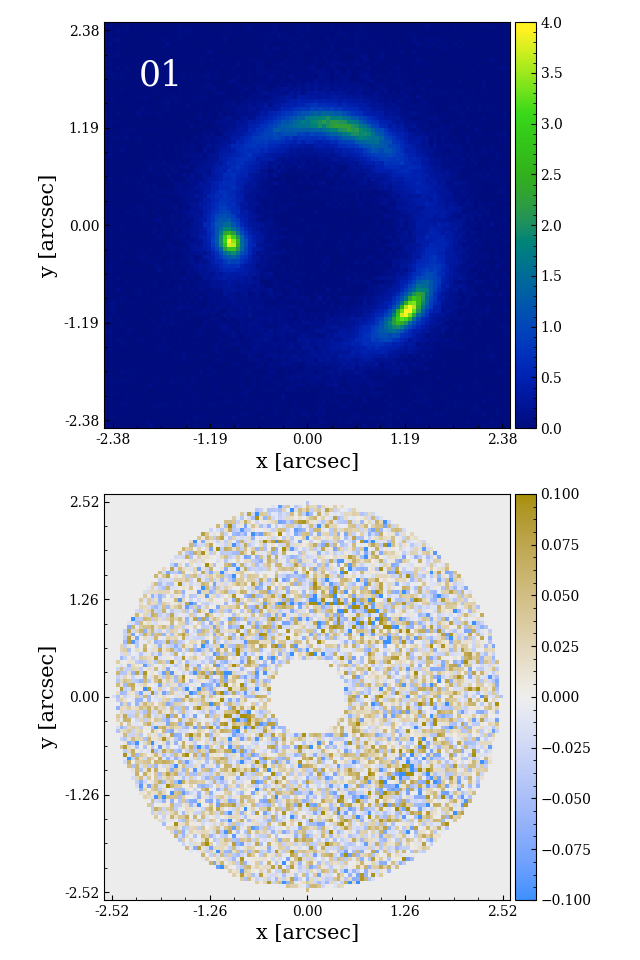}
        \caption{Upper panel: the mock image of the first system.
          Lower panel: the corresponding residuals after fitting a
          smooth macro model.}
    \label{fig:mock}
\end{figure}

\subsection{Lensing simulations}\label{sec:mocks}
The strong lens simulations used in this work represent the lensing
system with three components, the lens galaxy, the source galaxy, and
the line-of-sight dark matter haloes. 

\subsubsection{Lens and source}
We simulate the smooth mass distribution of our lens galaxy as a 2D singular
isothermal ellipsoid (SIE) of the form, 
\begin{equation}
    \begin{array}{lll}
        \Sigma(x, y) & = & \frac{\sigma^2}{2{\rm G}}\frac{1}{\sqrt{x^2 q + y^2/q}} \\
        & = & \frac{{\rm c}^2 }{8 \mathrm{\uppi G}}\frac{D_{\rm A} \left(0,\ z_{\rm source}\right)}{D_{\rm A} \left( z_{\rm lens},\ z_{\rm source}\right) D_{\rm A}\left(0,\ z_{\rm lens}\right)} \frac{R_{\rm E}}{\sqrt{x^2 q + y^2/q}}
    \end{array}, 
\end{equation}
where $R_{\rm E}$ is the projected Einstein radius, which can be related to the
velocity dispersion of the profile, $\sigma$, given the lens and
source redshifts ($z_{\rm lens}$ and $z_{\rm source}$, respectively) and a cosmology.
The quantity, $D_{\rm A}\left(z_1, z_2\right)$, is the angular
diameter distance between $z_1$ and $z_2$; $q$ is the axis ratio. For simplicity, we do not add any external shear to our mock observations. However, we include an external shear when modelling the lenses.

For the source, we adopt the \cite{Sersic1963} profile, 
\begin{equation}
    I(r) = I^{'}{\rm exp}\left[-b_{n}\left(\frac{r}{r_{\rm e}}\right)^{1/n}\right],
\end{equation}
where $I^{'}$ is the scale intensity, $n$ is the Sersic index and
$b_n$ is a function of $n$ defined such that $r_{\rm e}$ is the
half-light radius. However, the image residuals caused by
perturbations to the lens model (in our case the perturbing haloes)
are effectively a product of the deflection angles due to the
perturbation and the gradient of the source surface brightness
\citep{Vegetti2009}. The cuspy centre of the Sersic profile creates an
infinite surface brightness gradient which numerically causes infinite
surface brightness differences. This can be overcome by oversampling
the source light, but at the expense of increased computational run
time. Instead, we use a Sersic profile with a ``core'' \citep{Graham2003,
  Trujillo2004} to simulate the background source. This has the form,
\begin{equation}
    I(r) = I^{'}{\rm exp}\left[- b_n \left(\frac{r^\alpha + r^\alpha_{\rm c}}{r_{\rm e}^\alpha}\right)^{1/(n\alpha)}\right],
\end{equation}
where $r_c$ is the ``core radius'', which marks the transition radius
from the Sersic profile to constant surface brightness.  When
$r_c \to 0$, the profile reduces to the Sersic form. The parameter,
$\alpha$, quantifies how quickly the profile transitions from a
regular Sersic form to one with a constant surface brightness core. To
simplify both our simulation and modelling processes, we fix
$\alpha=2$ and $r_{\rm c}=0.01$". With these parameters fixed, the
source model has the same three free parameters as the regular Sersic
profile: $r_{\rm e}$, $n$ and $I^{'}$.

\subsubsection{Low-mass dark matter haloes}
Low-mass dark matter haloes that can perturb lensing arcs can be of
two types: subhaloes within the main lens, or central haloes at different
redshifts that happen to lie close to the path of  
light from the source galaxy to the observer. We will refer to the
latter as line-of-sight haloes.

The number and mass distribution of subhaloes within the lens are
somewhat uncertain. Subhaloes are subject to tidal stripping and
disruption; these effects are significantly enhanced in hydrodynamic
simulations compared to dark matter only simulations due to the
presence of the dense stellar component at the centre of the main lens
\citep{Sawala2017,Garrison-Kimmel2017,Richings2021}. The degree of
disruption also depends on the details of the galaxy formation model
\citep{Richings2020}.  In contrast, line-of-sight haloes are not
subject to these environmental effects and, in the mass range of
interest here, they are entirely `dark', having never formed stars.

Recent studies \citep{Li2016, Li2017, Despali2017, Caganseng2020} have
shown that for the lens and source redshifts typical of SLACS lenses \citep{Bolton2006},
the lensing perturbations mainly arise from line-of-sight haloes. This
is fortunate because the irrelevance of uncertain baryon effects,
makes line-of-sight haloes a particularly clean probe of a cutoff
scale in the halo mass function. In this work, we focus on these
low-mass, line-of-sight haloes whose mass function has the form given
by \citet{Lovell2014}:
\begin{equation}\label{equ:mass_function}
    F(M_{200}, z)\equiv\frac{d^{2}N}{dM_{200}dV}(M_{200}, z) = F_{\rm
      CDM}(M_{200}, z)\left(1 + \frac{M_{\rm
          hf}}{M_{200}}\right)^{-1.3}. 
\end{equation}
This is composed of two parts, the CDM mass function,
$F_{\rm CDM}\left(M_{200}, z\right)$, for which we use the form derived
by \citet{Sheth2001}, and a cutoff parameterized by the
`half-mode' mass, $M_{\rm hf}$, a characteristic mass corresponding to
the mass scale at which the dark matter transfer function falls to half the CDM
transfer function. The half-mode mass can be related to the mass of the dark matter
particle, $m_{\rm DM}$, as, 
\begin{equation}\label{equ:mp}
    M_{\rm hf} = 10^{10}\left(\frac{m_{\rm DM}}{{\rm keV}}\right)^{-3.33} {\rm M}_{\odot} {\rm h}^{-1},
\end{equation}
where $\rm{h}$ is the Hubble parameter in units of 100 km/s/Mpc
\citep{Lovell2014}. From the perspective of strong lensing,
$M_{\rm hf}$ may be regarded as an effective `cutoff' mass in the dark
matter mass function, below which there are very few haloes.

We draw low-mass dark haloes in a lightcone between the observer and
the source galaxy within an angular radius, $r_{\rm lc}$, chosen as,
\begin{equation}
    r_{\rm lc} = \left\{\begin{array}{ll}
         5.0^{\dprime} & \left(z \leq z_{\rm lens}\right) \\
         5.0^{\dprime} - 2.0^{\dprime} \times \frac{D_{\rm A}\left(z_{\rm lens},\ z\right)D_{\rm A}\left(0,\ z_{\rm source}\right)}{D_{A}\left(0,\ z\right)D_{\rm A}\left(z_{\rm lens},\ z_{\rm source}\right)} & \left(z > z_{\rm lens}\right)
    \end{array} \right.,
\end{equation}
where $D_A$ is the angular-diameter distance and $z_{\rm lens}$ and
$z_{\rm source}$ are the redshifts of the lens and source galaxies
respectively. The angular radius of the lightcone is fixed at
5.0$^{\dprime}$ (this is $\sim$ 3 times the Einstein radius, and
including haloes at larger radii hardly affects our lensing images) in
front of the main lens and gradually decreases behind from
5.0$^{\dprime}$ to 3.0$^{\dprime}$ at the source redshift, which
ensures that all perturbers that might influence the lensed image of
the source are included. The lightcone is evenly divided into redshift
bins, of width $\Delta z= 0.01$, which balances the accuracy and
computational cost of the calculation (see appendix \ref{AppendixA}
for details). Within each bin, we sample from the halo mass function,
assign density profiles from the halo mass-concentration relation
corresponding to that redshift, and place these perturbing haloes
randomly on the central redshift plane.  Neglecting the clustering of
the line-of-sight haloes is a good approximation if, over the angular
extent of interest, the typical line-of-sight separation of pairs of
haloes is much larger than their correlation length.  This is expected
to be true because: (a) the intrinsic clustering of low mass haloes is
weak \citep{Kaiser1984, Sheth1999}, and (b) the angular extent of
interest is very narrow since it is set by the width of the Einstein
ring image.

We generate low-mass haloes with $M_{200}$ between $10^6$ M$_\odot$
and $10^{10}$ M$_\odot$ within the light cone. Depending on the dark
matter model, a significant fraction of the mass in the Universe can
exist as collapsed objects in this mass range. To solve correctly the
strong lensing equations (which have already assumed a background
universe) we add negative mass sheets onto each plane to compensate
for the otherwise `double-counted' mass within perturbers in the
lightcone \citep{Petkova2014, Birrer2017b, Gilman2019}. For each plane, we include
a mass sheet with constant negative surface density equal to the
expected density in low-mass haloes on that plane. This leads to a
mass sheet with a surface density
\begin{equation}
    \Sigma_{\rm negative} = - \frac{1}{A}\int F(M_{200}, z) M(M_{200}, z, s) ~{\mathrm d}M_{200}{\mathrm d}s~{\mathrm d}V,
\end{equation}
where $A$ is the physical area of the lightcone at the plane's
redshift, $F(M_{200}, z)$ is the mass function and $M(M_{200}, z, s)$
is the total mass of a dark halo on the plane, which can differ from
$M_{200}$ due to halo truncation. The ``s'' denotes some other
parameters of the dark matter profile; in our case, as
we discuss below, it represents the ``concentration'' and
``truncation radius''.

Instead of modelling haloes as Navarro-Frenk-White (NFW) profiles
\citep{NFW1996, NFW1997}, which have an infinite total mass (when
integrating to infinite radius), we use a truncated NFW profile
\citep{Baltz2009},
\begin{equation}
    \rho(r) = \frac{M_0}{4{\rm \pi}r\left(r_{\rm s} + r\right)^2}\cdot\frac{r_{\rm t}^2}{r_{\rm t}^2 + r^2},
\end{equation}
which has an additional truncation term compared with the standard NFW
form. Here, the scale mass, $M_0$, and scale radius, $r_{\rm s}$, are a
pair of parameters that specify an NFW profile and $r_{\rm t}$ is the
truncation radius. With the addition of the truncation term,
$\rho(r) \propto r^{-5}$ at large radii, and the profile now has a finite
mass given by,
\begin{equation}\label{equ:m_tot}
    M_{\rm t} = M_{0}\frac{\tau^2}{\left(\tau^2 + 1\right)^2}\left[\left(\tau^2 - 1\right){\rm ln}\tau + \tau{\rm \pi} - \left(\tau^2 + 1\right)\right],
\end{equation}
where $\tau=r_{\rm t}/r_{\rm s}$. We set the truncation radius to be
the corresponding NFW $r_{100}$, the radius within which the average
density of an NFW profile is 100 times the critical density of the
universe. With $\tau$ fixed to $r_{100}/r_{\rm s}$, the tNFW profiles are
determined by the same two parameters as the NFW profile, $M_{200}$
and the concentration, $c = r_{200} / r_{\rm s}$, which quantifies how
centrally concentrated a halo is. Haloes that form earlier typically
have higher concentrations \citep{NFW1996,NFW1997}. Since the
reduction in small-scale power in a WDM model leads to later formation
time for low mass haloes these are less concentrated than their CDM
counterparts \citep{Lovell2012,Schneider2012, Maccio2013, Bose2016,
  Ludlow2016}.

When considering WDM models we adopt the mass-concentration relation
of \citet{Bose2016}
\begin{equation}\label{equ:mc}
    c(M_{200}, z) = c_{\rm CDM}(M_{200}, z)\left(\left(1 + z\right)^{0.026z - 0.04}\left(1 + 60\frac{M_{\rm hf}}{M_{200}}\right)^{-0.17}\right),
\end{equation}
where $c_{\rm CDM}(m, z)$ is the mass-concentration relation in the
CDM case. For CDM, we use the mass-concentration relation of
\citet{Ludlow2016}, which agrees with recent simulation results from
\citet{Wang2020}, who measured halo concentrations in $\Lambda$CDM
simulations over twenty orders of magnitude in halo mass. We use the
public package \textsc{COLOSSUS} \citep{Diemer2018} to compute the
mass-concentration relation. The \citet{Ludlow2016} relation is for
the median concentration at a given mass, but simulations predict
scatter in concentration at fixed mass which is well described by a
log-normal distribution \citep{Neto2007}. As a result, we include
log-normal scatter in the mass-concentration relation, with a standard
deviation of 0.15 dex, as shown by \citet{Wang2020}.

Multi-plane
ray tracing is necessary to calculate the effect of line-of-sight
haloes on flux ratio anomalies with sufficient precision to match observations \citep{Gilman2019}. We use the publicly available software package \textsc{PyAutolens}
\citep{Nightingale2015, Nightngale2018, Nightingale2019, pyautolens} to simulate
lenses using the light and mass profiles described above. To define
our fiducial lens sample, we simulate 50 unique lensing systems with
different values for the lens SIE mass model (e.g. Einstein radius,
axis ratio, etc.) and source core-Sersic profile. The input
$M_{\rm hf}$ is 10$^7$ M$_\odot$. For all 50 systems we assume lens
redshifts of 0.5 and source redshifts of 1. 

We produce observations representative of Hubble Space Telescope
imaging, with a pixel size of $0.05\arcsec$, a Gaussian PSF with a
standard deviation of $0.05\arcsec$ and background sky level of
1e$^{-}$/pix/second. We set the exposure time to be 600\,s. The
source surface brightness of each simulation is chosen such that the
maximum signal-to-noise ratio (e.g.\ in the brightest pixel) has a
value $\sim50$. Table~\ref{tab:drawn_rges} summarizes the models
and parameters for our mock observations, where closed brackets
indicate the range of values from which each parameter is randomly
drawn. For convenience, we use $\{T\}_{\rm fiducial}$ to denote the
set of true parameter values used for our fiducial tests and a
superscript $i$ to represent the parameters of the $i$-th lensing
system. Mock images of our 50 strong lensing systems are shown in
Fig.~\ref{fig:arcs}.

\subsection{Analysis Process}
Our analysis of each mock observation consists of two steps: smooth
model fitting followed by ABC inference. Below we describe the
detailed steps in this process and illustrate them for one system.

\subsubsection{Smooth Model Fitting}
First, we fit a simple lens model to the mock lensing image. This
consists of a smooth parametric model plus an external shear for the lens galaxy mass distribution (the {\it macro model}) and a smooth parametric model for the source. We adopt the same parametric forms we used when generating
the simulated mock image, an SIE lens mass model and a core-Sersic source.

We use \textsc{PyAutoLens} to fit the mock images, and adopt
\textsc{PyMultiNest} with constant sampling efficiency mode switched
on \citep{Feroz2009, Buchner2014} for the non-linear
optimisation. Table~\ref{tab:mock_paras} shows the inferred parameters
for this fit, alongside the true input values for the first mock
image. To distinguish between the true inputs and the best-fit values
we label the latter as $\{F\}_{\rm fiducial}$, where a superscript,
$i$, denotes the best-fit parameters for the $i$-th lensing
system. This table shows that the inferred and input lens model are
close, but not in perfect agreement; this is due to the effects of the
low-mass dark matter haloes. We will show later that this small
mismatch does not affect our calculation.

In the upper and lower panels of Fig.~\ref{fig:mock} we show the
``observed'' image and best-fit residuals for one of our 50 mock
observations. To develop intuition on how the low-mass perturbers
affect the image residuals, we further show the ``effective
convergence" (the divergence of deflection angles) of low mass
perturbers and the corresponding best-fit residuals in
Fig.~\ref{fig:eff_cvgc}. We assume a very small constant noise to show
the residuals clearly. The three strong lensing systems shown have the
same lens and source galaxy as in Fig.~\ref{fig:mock}, except that the
source for the system of the right-hand column is at $z$ = 2.5. For a clear
comparison, we make sure that the three systems share the same
realisation of low mass purturbers. To achieve this, instead of
applying the mass function described by $M_{\rm hf}$ in
Eq.~\ref{equ:mass_function}, we draw haloes from the mass function of
\citet{Sheth2001} but discard those with $M_{200}$ below a certain
value. Furthermore, we assume no correlation between $M_{\rm hf}$ and
the mass-concentration relation. Please note that this change of low-mass perturber realisation is only implemented for
Fig.~\ref{fig:eff_cvgc}; in the rest of this paper, we  use
the mass function and mass-concentration relation described by
Eq.~\ref{equ:mass_function} and Eq.~\ref{equ:mc} respectively.

In Fig.~\ref{fig:eff_cvgc}, the systems in the left and right columns
have a cutoff in the mass function at $10^7$ M$_\odot$, while the
system in the middle column has a cutoff at $10^9$ M$_\odot$. As seen
from the left two columns, the system with a lower cutoff in the mass
function has far more low-mass perturbers and thus more subtle
structure in the residuals. When comparing the left and the right
columns, we see that the number of low-mass perturbers increases
significantly with the source redshift and the ``profiles'' of
individual perturbers are more heavily distorted because of the
multi-plane lensing effect. Also, for a higher redshift source galaxy,
the residuals are larger.

\subsubsection{ABC inference through forward modelling}

Approximate Bayesian Computation (ABC) is a likelihood-free method
suitable for problems where the likelihood is difficult to express
analytically, but model predictions are relatively easy to
simulate. It has been widely applied in astrophysics, for example in
studies of large-scale structure, planet surveys and reionization
\citep{Akeret2015, Hahn2017, Davies2018, Hsu2020}. \citet{Birrer2017},
\citet{Gilman2019, Gilman2020a, Gilman2020b} and \citet{Enzi2020} have
also used ABC to constrain dark matter substructure using strong
lensing data. To begin, one defines a summary statistic which measures
the similarity between the observational data and the simulations.  In
principle, any statistic (as long as it contains information of
interest) can be used as a measure of similarity, but different
statistics may have different requirements on data quality, sampling
efficiency, etc. A good statistic captures as much of the
characteristic information of different models as possible: in our
case, the lensing perturbations from low-mass dark matter
haloes. Finding a proper summary statistic is a challenging part of
ABC and it is beyond the scope of this work to investigate the many
possible candidate summary statistics. Instead, following
\citet{Bayer2018}, we use the power spectrum of the best-fit image
residuals to construct a summary statistic that we hope can extract
information on the number of low-mass dark matter haloes perturbing
the lens.
 
\begin{figure}
	\includegraphics[width=0.99\columnwidth]{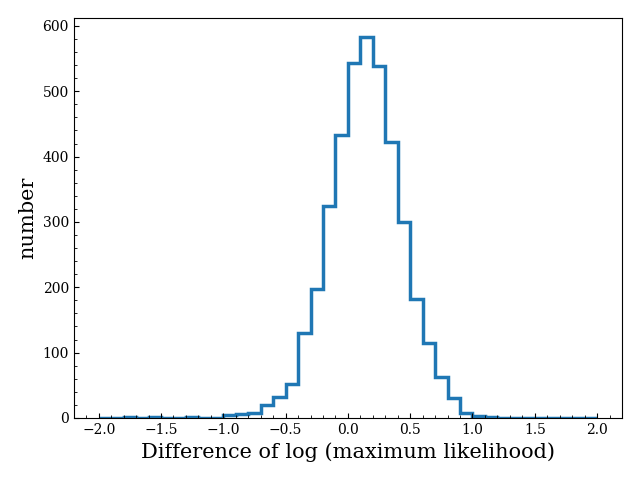}
        \caption{Histogram of the differences between the log maximum likelihood obtained with the priors listed in Table~\ref{tab:fit_priors} and that obtained with 5 times larger priors.} 
    \label{fig:lklhd_diff}
\end{figure}

Based on the best-fit source and macro model for the lens,
$\{ F \}^1_{\rm fiducial}$, we simulate images of this best-fit
source and lens combination with the addition of low-mass haloes along
the line-of-sight. We uniformly sample
$\log_{10} M_{\rm hf} / \rm M_\odot$ between 7 and 10 (corresponding
to assuming a flat prior on $\log_{10} M_{\rm hf}$ between
$10^{7}$~M$_\odot$ and $10^{10}$~M$_\odot$) and, for each sampled
$M_{\rm hf}$ value, we draw a random realisation of low-mass dark
matter haloes with the appropriate mass function and
mass-concentration relation. We do this for 20,000 values of
$M_{\rm hf}$, producing a corresponding lensed image in each case and
then fit each of these 20,000 images to obtain the best-fit image
residuals. We have confirmed that our results are converged with
20,000 samples. Note that the forward-modelled images are simulated
from the known lens and source galaxies, so when performing the
fitting, we add narrow priors centred on the parameter values of the
known best-fit lens and source models to accelerate
the fitting processes. Since our only goal is to find the maximum
likelihood model, once our priors contain the correct solution, the
size of the priors does not affect our results. In Table~\ref{tab:fit_priors}, we summarize the type and size of the priors we used for the fitting processes. To make sure the priors we take are broad enough to find the maximum likelihood model, we fit 4000 different realisations with both the listed priors and 5 times larger priors and then compare the maximum likelihoods obtained. Here, ``5 times larger priors'' means that for the priors listed in Table~\ref{tab:fit_priors}, if it is a Gaussian prior, then the sigma is taken 5 times larger than the value listed. In Fig.~\ref{fig:lklhd_diff}, we show the histogram of the difference of the log maximum likelihood. As shown, the log maximum likelihood results of larger and smaller priors are very close with only $\sim$ 1.0 difference, which verifies that the priors we take are broad enough to contain the maximum likelihood model.

\begin{table}
    \renewcommand\arraystretch{1.5}
	\centering
	\begin{tabular}{cc} 
		\hline
		& Prior \\
		\hline
		\textbf{Lens} & Elliptical Isothermal \\
		centre (x, y) [$^{\dprime}$] & G$\left({\rm input}, 0.01\right)$ \\
		$R_{\rm E}$ [$^{\dprime}$] & G$\left({\rm input}, 0.008\right)$ \\
		axis ratio & G$\left({\rm input}, 0.03\right)$ \\
		position angle [$^\circ$] & G$\left({\rm input}, 2.0\right)$\\
		\hline
		\textbf{External Shear} & \\
		magnitude & G$\left(0.0, 0.05\right)$ \\
		position angle [$^\circ$] & U$\left(0.0, 180.0\right)$ \\ 
		\hline
		\textbf{Source} & Elliptical Core-Sersic \\
		centre (x, y) [$^{\dprime}$] & G$\left({\rm input}, 0.01\right)$ \\
		$r_{\rm e}$ [$^{\dprime}$] & G$\left({\rm input}, 0.015 \times {\rm input}\right)$ \\
		$n$ & G$\left({\rm input}, 0.1\right)$ \\
		$I^{'}$ [e$^{-}$ pix$^{-1}$ s$^{-1}$] & G$\left({\rm input}, 0.16\times{\rm input}\right)$ \\
		axis ratio & G$\left({\rm input}, 0.02\right)$ \\
		position angle [$^\circ$] & G$\left({\rm input}, 2.0\right)$ \\
		\hline
	\end{tabular}
	\caption{Priors used in fitting processes. G$\left(a, b\right)$ represents a Gaussian prior centering on $a$ with standard deviation $b$. U$\left(a, b\right)$ represents a uniform prior between $a$ and $b$. The ``input'' here refers to the corresponding input value used when simulating the mock image.}\label{tab:fit_priors}
\end{table}

\begin{figure}
	\includegraphics[width=0.99\columnwidth]{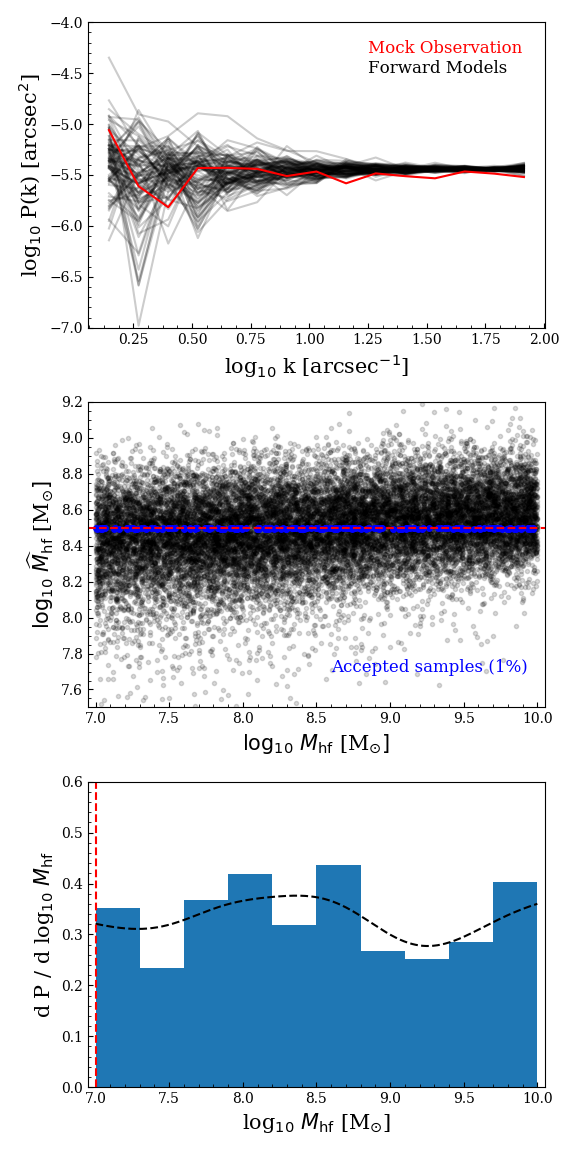}
        \caption{\textbf{Upper panel}: the power spectrum computed
          from the best-fit image residuals. The red line is the power
          spectrum of the mock observation. Black lines are for 100
          forward models. \textbf{Middle panel}: the summary
          statistic, ${\rm log}_{10} \widehat{M}_{\rm hf}$, 
          computed from the power spectra. The black points are
          ${\rm log}_{10} M_{\rm hf}$ for the forward models. The horizontal dashed red
          line marks the value for the mock observation and
          points 1\% closest to it are marked in blue. \textbf{Lower
            panel:} The posterior distribution of $M_{\rm hf}$, formed
          by collecting together the $M_{\rm hf}$ of the blue points is
          shown in the middle panel. The vertical dashed red line
          marks the true input of $M_{\rm hf}$. The black dashed curve
          is a kernel density estimate corresponding to the
          histogram.}
    \label{fig:pws}
\end{figure}

Next, for the images of the best-fit residuals, we set pixels outside
the annular region between $0.5^{\dprime}$ and $2.4^{\dprime}$ to be 0
and then Fourier transform the residual image and azimuthally average
to obtain the 1-D power spectrum, $P(k)$. In the upper panel of
Fig.~\ref{fig:pws}, we show power spectra for the forward-modelled
residuals as black lines (for clarity we only plot 100 of them); the
red curve marks the power spectrum of the image residuals of the
original observation. Each line is composed of 15 $P(k_i)$ values with
the $k_i$ logarithmically spaced between $\sim$ 1.5 arcsec$^{-1}$ and
85 arcsec$^{-1}$ ($\sim$ 2${\rm \uppi}$ / (4.2 arcsec) to
2${\rm \uppi}$ / (0.07 arcsec)).
 
 Although we have converted the images into power spectra, we cannot
 simply use the 1D ``power spectrum" as our statistic. The reason is
 that each power spectrum is composed of 15 different values and for
 ABC to converge at a point in a 15-dimensional space, a very small
 acceptance rate must be set, which, in turn, requires far more forward
 models. To further reduce the ``dimensionality'' of the statistic,
 following \citet{Fearnhead2012}, we generate a summary statistic from
 a linear combination of the logarithm of the power spectrum values at
 different $k$, 
 \begin{equation}
     {\rm
   log}_{10} \widehat{M}_{\rm hf} = \sum_{i=1}^{15} \beta_{i} \log_{10} \left[ P(k_i) /
       \mathrm{arcsec}^2 \right] + \beta_{0}.
\label{eqn:shat}
 \end{equation}

 We use ${\rm log}_{10} \widehat{M}_{\rm hf}$ as the summary statistic
 for the ABC inference. The coefficients $\{\beta_{i}\}$ (including
 $\beta_0$) are obtained by minimizing the differences between the
 input ${\rm log}_{10} M_{\rm hf}$ values used in the forward models
 and their corresponding ${\rm log}_{10} \widehat{M}_{\rm
   hf}$. Specifically, we find the values of $\{\beta_{i}\}$ that
 minimise,
 \begin{equation}
     \sum_{j=1}^{20000}\left\|{\rm log}_{10} \widehat{M}_{\rm hf}^{(j)} - {\rm log}_{10} M^{(j)}_{\rm hf}\right\|^2,
 \end{equation}
 where ${\rm log}_{10} \widehat{M}_{\rm hf}^{(j)}$ is computed from
 the $j$-th forward-modelled power spectrum and $M^{(j)}_{\rm hf}$ is
 the corresponding input half-mode mass. Since the forward-modelled
 power spectrum is systematically different for different macro
 lensing configurations, we optimize $\{\beta_{i}\}$ separately for
 each of our 50 lensing systems. In the middle panel of
 Fig.~\ref{fig:pws}, we plot the value of
 ${\rm log}_{10} \widehat{M}_{\rm hf}$ for each forward-modelled
 simulation as a black point. The horizontal dashed red line marks the
 value of ${\rm log}_{10} \widehat{M}_{\rm hf}$ given by the mock
 observation.
 
 We accept the forward-modelled simulations with the 1\% of ${\rm log}_{10} \widehat{M}_{\rm hf}$
 values closest to the observed ${\rm log}_{10} M_{\rm hf}$ (Eqn.~{\ref{eqn:shat}}),
 which are shown as blue points in the middle panel of
 Fig.~\ref{fig:pws}. The set of $M_{\rm hf}$ values associated with
 those blue points are then a sample drawn from the posterior
 distribution of $M_{\rm hf}$ (following the ABC method); their
 density can be used to estimate the posterior density of
 $M_{\rm hf}$. 

 In the lower panel of Fig.~\ref{fig:pws} we plot the posterior
 density for $\log_{10} M_{\rm hf}$, where the vertical dashed red
 line shows the true input, $M_{\rm hf} = 10^7$ M$_\odot$. It is clear
 that with just one system and the data quality of our fiducial setup,
 we are not able to derive a tight constraint on $M_{\rm hf}$. While
 improved data quality might improve this somewhat, a single lens
 system will always be limited by the stochastic nature of the
 distribution of the low-mass dark matter haloes we are trying to
 detect. Just because a particular value of $M_{\rm hf}$ allows
 perturbers of a given mass this does not mean that there will happen
 to be one in a location where it produces detectable image
 residuals. As such, tight constraints will rely on combining results
 from multiple systems.

 \begin{figure*}
	\includegraphics[width=2.0\columnwidth]{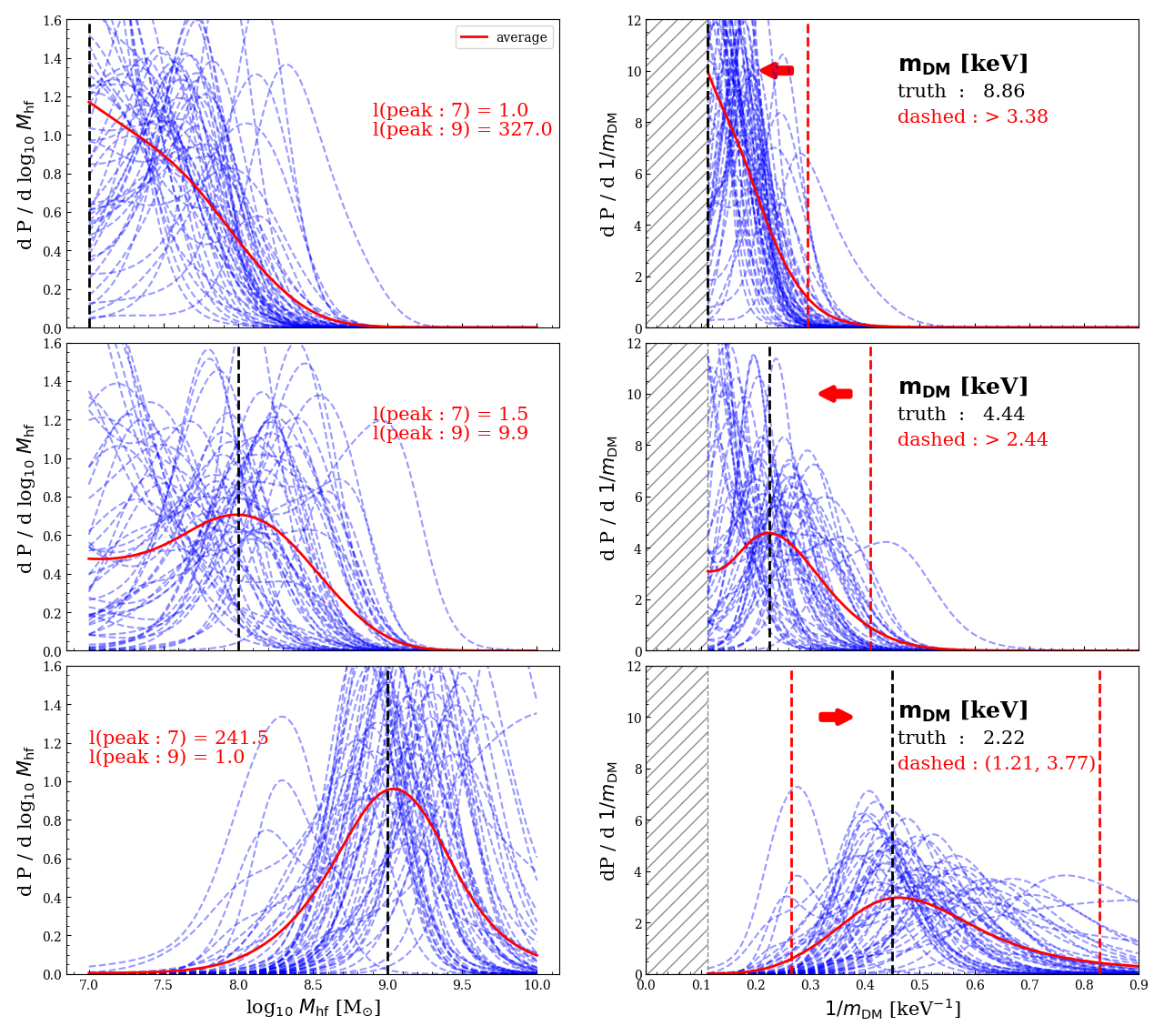}
        \caption{Tests on the fiducial setting. The left column shows
          the constraints on $M_{\rm hf}$ (adopting a flat prior on
          $\log M_{\rm hf}$) and the right columns the constraints on
          $1/m_{\rm DM}$ (adopting a flat prior on $1/m_{\rm
            DM}$). From top to bottom, panels correspond to true
          $M_{\rm hf}$ values of $10^7$, $10^8$, $10^9$ M$_\odot$,
          marked by the black vertical dashed lines. Blue dashed
          curves are individual constraints for each set of 50 lensing
          systems (for clarity, we only plot 50 out of 500 here). The
          red curve is the average constraint from the 500 sets. In
          the right column we use the mean posteriors to place 2$\sigma$(95\%)
          credible interval limits (upper limits on $1/m_{\rm DM}$ for
          the top 2 panels and a both upper and lower limit for the bottom one). The
          hatched regions are where $M_{\rm hf}$ < $10^7$ M$_\odot$,
          which is outside the range probed by our forward models.}
    \label{fig:accuracy}
\end{figure*}

\section{Results and Discussion}\label{sec:results}
In this Section, we will first show that the forward modelling
procedure described above can correctly recover the input value of
$M_{\rm hf}$ when combining results from multiple lenses. Then we will
explore how the precision of the constraints depends upon the lensing
configuration and image quality of the lenses in an observed
sample. We will then compare the strength of our method with the
method based on quasar flux ratio anomalies.

\subsection{Tests of the accuracy of the method}

By repeating the procedure described above for each observed
system we can obtain a constraint from our mock observations of 50
lensing systems.  Just as for a real set of 50 observed systems, the
posterior probability density for $M_{\rm hf}$ will not necessarily peak at the
true value. To assess whether our method is biased, we create
500 sets of 50 observed lenses. Running the full procedure described
above 500 times, each time for 50 different lenses, would be
prohibitively expensive because of the very large number of forward
models this would require. We therefore use the same set of 50 macro
lens and source model parameters ($\{ T \}_{\rm fiducial}$) for each
of our 500 sets, only changing the realisations of low-mass dark
matter haloes between the sets of 50 observed images. This allows us
to reuse the same forward models for each of our 500 sets.

In principle, we should generate a new set of forward models for each
of the 500 sets (together 25000 lenses) because the particular
realisation of the dark matter haloes affects the best-fit macro model,
which is the model used to generate the forward model
images. 
We will see later that even though there is a slight mismatch in
best-fit models for different realisations, we can still correctly
recover the true $M_{\rm hf}$. Similarly, to examine how the method
behaves with different input (true) values for $M_{\rm hf}$, we
further simulate 500 sets of 50 observations with $M_{\rm hf}=10^8$
and $10^9$ M$_\odot$ and apply ABC inference to them, all using the
same forward models previously simulated. 

Note that some care needs to be taken when combining results from
multiple ABC calculations. This is because ABC produces an estimate of
the posterior distribution, rather than the likelihood. Combining
multiple measurements in a traditional Bayesian analysis (with a
calculable likelihood function) is a case of multiplying the
likelihoods of the different measurements together to get a total
likelihood and then multiplying this by the prior in order to find
(something proportional to) the posterior distribution. In this work
we use a flat prior on ${\rm log_{10}}\ M_{\rm hf}$, and thus the
posterior density (per unit ${\rm log_{10}}\ M_{\rm hf}$) we obtain is
proportional to the likelihood. This means that the posterior
densities per unit ${\rm log_{10}}\ M_{\rm hf}$ for individual systems
(such as the blue histograms in the lower panel of Fig.~\ref{fig:pws})
can be multiplied together to produce something proportional to the
joint likelihood, which can then be multiplied by a prior (once) to
get the posterior.

Before multiplying individual constraints, to reduce the noise and
obtain a smooth likelihood, we apply a kernel density estimation
method to the distribution of ${\rm log_{10}}\ M_{\rm hf}$ with a
kernel width of 0.3\,dex. We have tested smaller kernel sizes, but the
constraints quickly become noisy without significant improvement. To
correct for boundary effects, we choose the ``renormalization''
correction for our kernel density estimations.\footnote{We use the
  public package \textbf{PyQt-Fit} to implement the kernel density
  estimation (see ``https://pythonhosted.org/PyQt-Fit/'').} As an
example, the kernel density estimation for the fiducial result is
shown as the black dashed curve in the lower panel of
Fig.~\ref{fig:pws}.

The left column of Fig.~\ref{fig:accuracy} shows our test results for
three different cases with true input $M_{\rm hf}$ of $10^7$
M$_\odot$, $10^8$ M$_\odot$ and $10^9$ M$_\odot$ from top to bottom,
respectively. Each blue line is a constraint from 50 systems (for
clarity, we only show 50 blue lines in the plots). Any one set of 50
observations will not necessarily have a posterior that peaks at the
true value, so in order to assess whether our results are
systematically biased -- as opposed to just subject to random error --
we plot a mean posterior distribution from the 500 sets of 50
observations. These are the red lines in Fig.~\ref{fig:accuracy}; they
peak close to the input values of $M_{\rm hf}$, suggesting that our
method is unbiased. Using these mean posteriors we compute the ratio
of the posterior between the peak value and the values at
$M_{\rm hf}=10^7$ M$_\odot$ and $M_{\rm hf}=10^9$ M$_\odot$. With an
input value of $M_{\rm hf}=10^7$ M$_\odot$, the recovered constraint
shows that the model with $M_{\rm hf}=10^9$ M$_\odot$ is disfavoured
with a posterior density $\sim$ 327 times smaller than that of the
peak. In a WDM universe (e.g. with $M_{\rm hf}=10^9$ M$_\odot$),
$M_{\rm hf}=10^7$ M$_\odot$ would be ruled out, with a mean posterior density $\sim$ 242 times smaller than at the peak.

In the right column of Fig.~\ref{fig:accuracy}, we show our posterior
distribution in terms of $1/m_{\rm DM}$, which is the way in which
constraints on the DM particle mass from the Lyman-$\alpha$ forest are
typically expressed \citep{Irsic2017}. Note that the $1/m_{\rm DM}$
posteriors are not simply the $M_{\rm hf}$ posteriors transformed to a
new parameterisation. Instead, we transform the likelihood as a
function of $M_{\rm hf}$ to the likelihood as a function of
$1/m_{\rm DM}$ following Eq.~\ref{equ:mp}, and then adopt a flat prior
on $1/m_{\rm DM}$ (as done in Lyman-$\alpha$ studies). This is
different from a flat prior on $\log_{10} M_{\rm hf}$ and so the posteriors
are not actually the same in the two columns of
Fig.~\ref{fig:accuracy}.

Depending on the ``behaviour'' of the red curves, we place either a 2$\sigma$ (95\%) upper or lower limit on $1/m_{\rm DM}$. For the cases with true
$M_{\rm hf}=10^7$ and $10^8$ M$_\odot$, we place an upper limit, while
for the case with $M_{\rm hf}=10^9$ M$_\odot$ we place both lower and upper limits.
These limits are shown as vertical dashed red lines in the figure. We
can see that with 50 fiducial-like lensing systems, at 2$\sigma$
level, one can rule out particle candidates with $m_{\rm DM}$ less
than $3.38$ keV and $2.44$ keV in universes with true
$m_{\rm DM}=8.86$ and $4.44$ keV respectively, or rule out particles
with $m_{\rm DM} > 3.77$ keV and $m_{\rm DM} < 1.21$ keV in a universe with true $m_{\rm DM}=2.22$
keV. We view the 2$\sigma$ limits in the first two panels as
conservative because we do not assign any posterior mass to the shaded
regions in our computation. If we had simply assumed that the
posterior density in the shaded region was the same as at $\sim$ 0.12
keV$^{-1}$, the upper limits would have been tighter. 

\begin{figure*}
\includegraphics[width=2.0\columnwidth]{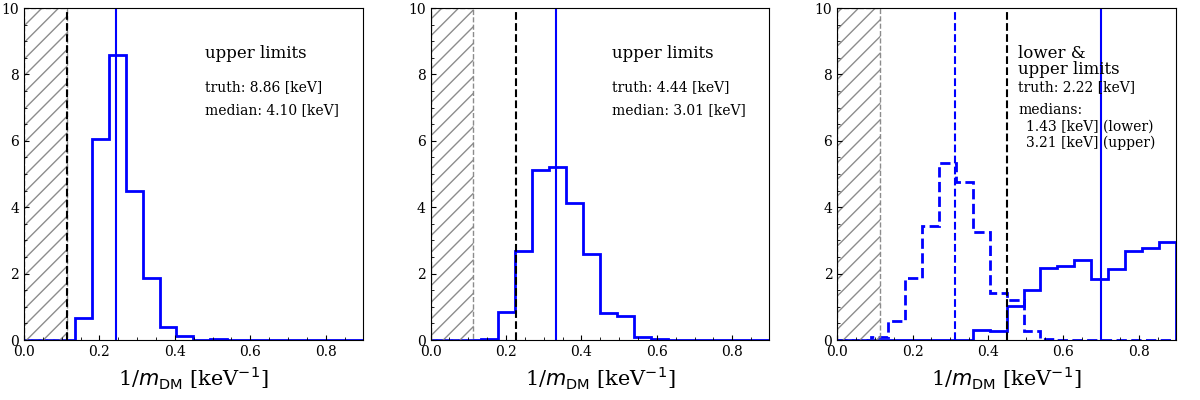}
\caption{Histograms of 2$\sigma$ (95\%) upper or both upper and lower limits on $1/m_{\rm DM}$. From left to right the input half-mode mass is $M_{\rm hf}=10^7, 10^8$ and $10^9$ M$_\odot$, with the corresponding DM particle masses written in the relevant panel. The limits are upper limits on $1/m_{\rm DM}$ (i.e. a lower limit on the DM particle mass), except for the case of $M_{\rm hf}= 10^9$ M$_\odot$ where both histograms of upper (solid line) and lower (dashed line) limits on $1/m_{\rm DM}$ are shown. The black dashed lines mark the true inputs. The vertical blue lines mark the medians of the histograms and corresponding median values are also listed.}
\label{fig:hist_2sigma}
\end{figure*}

Note that the constraints discussed above are average results. Observations of any 50 specific lensing systems would yield a constraint like one of the blue dashed lines in the figures, which might be tighter or looser than the average constraint. To demonstrate actual constraint one would expect to get with 50 observations, in Fig.~\ref{fig:hist_2sigma} we show the histograms of 2$\sigma$ limits from the 500 different realisations of a set of lensing observations (for true $m_{\rm DM}=8.86, 4.44$~keV, we compute lower 2$\sigma$ limits, and for true $m_{\rm DM}=2.22$~keV, we compute lower and upper 2$\sigma$ limits). As shown by the histograms, when true $m_{\rm DM}=8.86$~keV, the median 2$\sigma$ constraint we would get is $1/m_{\rm DM} < 0.24$~keV, and thus $m_{\rm DM} > 4.10$~keV. In other words, there is 50\% chance to constrain $m_{\rm DM}$ better than 4.10~keV with 50 lenses of similar settings. If true $m_{\rm DM}=4.44$~keV, the median 2$\sigma$ constraint is m$_{\rm DM} > 3.01$~keV. If true $m_{\rm DM}$=2.22~keV, the median constraint obtained is 1.43~keV$ < m_{\rm DM} < 3.21$~keV.

\begin{figure*}
	\includegraphics[width=2.0\columnwidth]{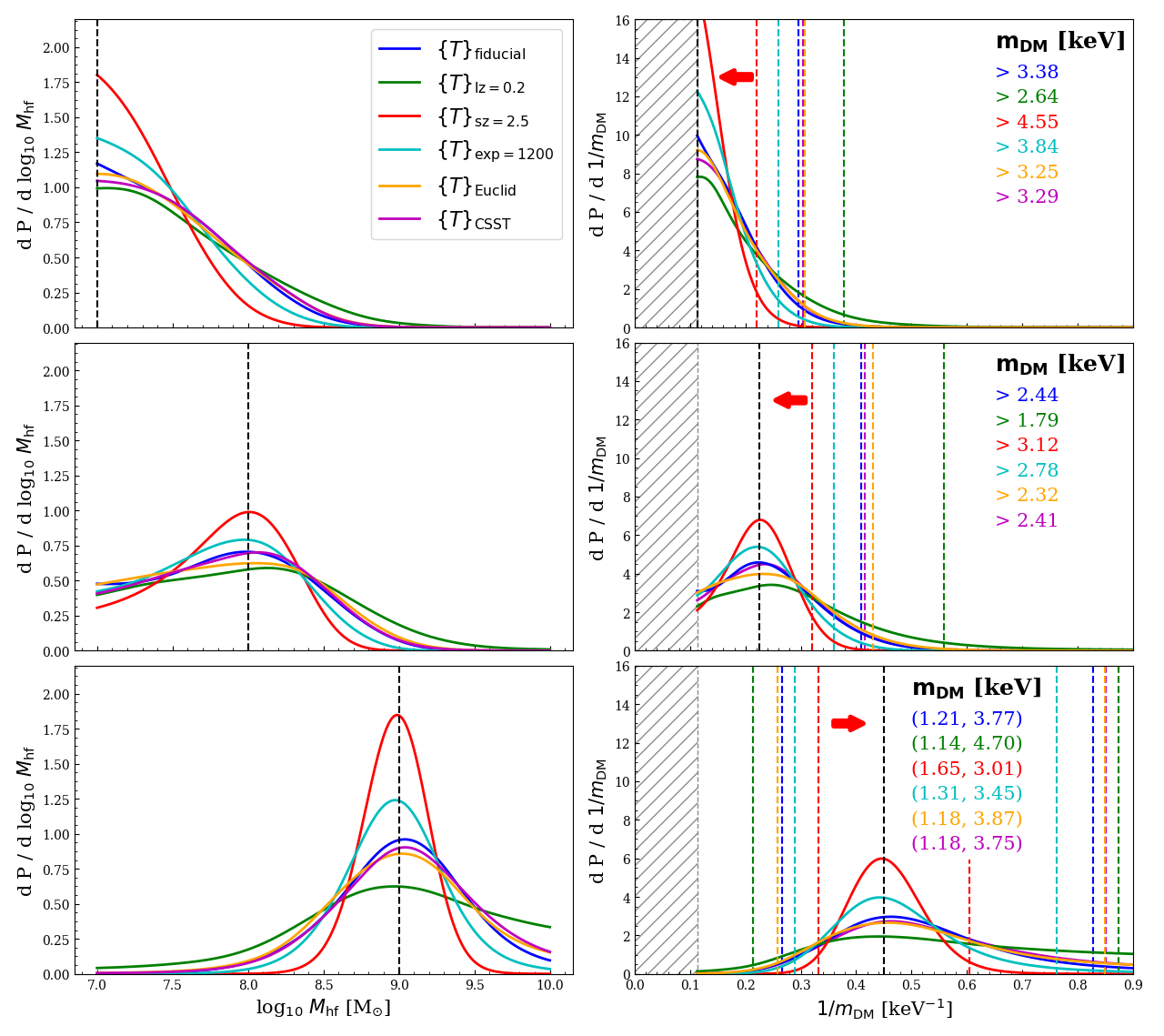}
        \caption{Average constraints from 50 lensing systems with
          different settings. As in Fig.~\ref{fig:accuracy}, the left
          panels show the constraints on $M_{\rm hf}$ and the right
          panels the constraints on $1/m_{\rm DM}$. Blue, green, red,
          cyan, orange and purple lines correspond to the results of
          settings, $\{T\}_{\rm fiducial}$, $\{T\}_{\rm lz=0.2}$,
          $\{T\}_{\rm sz=2.5}$, $\{T\}_{\rm exp=1200}$,
          $\{T\}_{\rm Euclid}$ and $\{T\}_{\rm CSST}$
          respectively. The vertical dashed black lines mark the true
          input values and the color lines the corresponding 2$\sigma$
          upper limits (top 2 panels) or both upper and lower limit (bottom panel) on
          $1/m_{\rm DM}$.}
    \label{fig:dependency}
\end{figure*}

\begin{figure*}
\includegraphics[width=2.0\columnwidth]{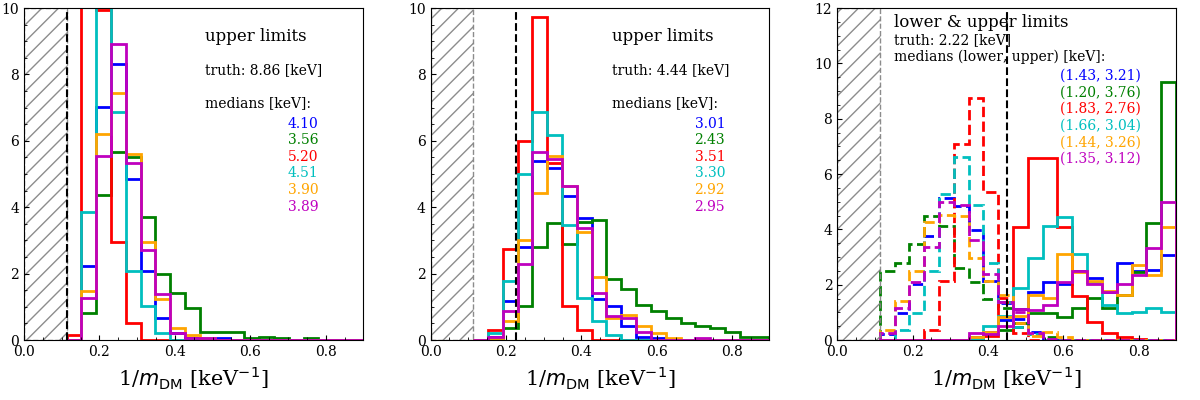}
\caption{Histograms of 2$\sigma$ (95\% upper) limits (solid lines) and lower limits (dahsed lines, only in the rightmost panel) on $1/m_{\rm DM}$. The vertical black dashed lines mark the true input $m_{\rm DM}$ in each panel. The colours of the histograms follow the meaning of in Fig.~\ref{fig:dependency}. The medians of the histograms are also listed in corresponding colors in the panels.}
\label{fig:hist_2sigma_multi}
\end{figure*}

\subsection{Dependency on the lensing configuration}

Having shown that with a large number of lensing systems the forward
modelling procedure can correctly recover the true $M_{\rm hf}$
(albeit with fairly broad posteriors), we now explore how the
constraints change when the properties of the lenses and sources are
varied. In particular, we vary the lens and source redshifts, the
image S/N ratio, and the image resolution.

To show clearly how the constraints depend upon a specific parameter,
we only change one parameter value at a time. For example, to
investigate how the result changes for a higher redshift source, we
only change the source redshift, and leave the image resolution,
Einstein radius, surface brightness, etc., unchanged from the
$\{T\}_{\rm fiducial}$ values. Notice that the parameters listed in
Table~\ref{tab:drawn_rges} are observational rather than intrinsic
physical quantities, so by fixing a parameter, we refer to fixing a
particular observational quantity. For example, keeping the source
surface brightness unchanged from a redshift of 1 to 2.5 means that
the source galaxy at $z=2.5$ is actually intrinsically brighter because of
cosmological dimming. The reason for changing quantities in this way
is that we want to focus on how the results change from an
observational perspective and thus provide a basic idea of what type
of lensing configuration has most constraining power, which may help
inform future observational designs.

In total, we have carried out five additional tests: placing the lens
at a lower redshift of 0.2, denoted as $\{T\}_{\rm lz=0.2}$; placing
the source at a higher redshift of 2.5, denoted as
$\{T\}_{\rm sz=2.5}$; doubling the exposure time, denoted as
$\{T\}_{\rm exp=1200}$; lowering the resolution, with  pixel size of
0.1$^{\dprime}$ and  PSF $\upsigma$ of 0.08$^{\dprime}$, similar to
the expected resolution of images from the Euclid Space Telescope
\citep{Collett2015}, denoted as $\{T\}_{\rm Euclid}$; assuming a
similar resolution to that of the China Space Station Telescope
(CSST), with pixel size of 0.075$^{\dprime}$ and PSF $\upsigma$ of
0.08$^{\dprime}$, denoted $\{T\}_{\rm CSST}$.
To first focus on the changes caused by different settings, without being
affected by set-of-observations to set-of-observations noise, we
compare the mean posteriors from 500 sets of 50 observed systems
(i.e. the equivalents of the red lines in Fig.~\ref{fig:accuracy}). 

In the left column of Fig.~\ref{fig:dependency} we plot the expected
$M_{\rm hf}$ posteriors for different lensing configuration using 
different colours. Blue, green, red, cyan, orange and purple
correspond to the settings, $\{T\}_{\rm fiducial}$, $\{T\}_{\rm
  lz=0.2}$, $\{T\}_{\rm sz=2.5}$, $\{T\}_{\rm exp=1200}$, $\{T\}_{\rm
  Euclid}$ and $\{T\}_{\rm CSST}$, respectively. Different rows show
results assuming different true inputs of $M_{\rm hf}$, which are
marked as vertical dashed black lines. As in Fig.~\ref{fig:accuracy}, 
we plot $1/m_{\rm DM}$ posteriors in the right column and place upper
(in the top two panels) or both lower and upper (in the bottom panel) 2$\sigma$ limits
on $1/m_{\rm DM}$, which are marked as vertical dashed lines. The red,
green and blue curves show how the results vary with the lens and source
redshifts.  

Increasing either the source or lens redshift improves the constraints
because (at fixed Einstein radius) the volume in which low-mass dark
matter haloes are projected close to the observed Einstein ring
increases as either redshift is increased. The comparison between the
cyan and blue curves shows that increasing the S/N ratio (exposure
time) results in a better constraint. Also, better angular resolution gives better constraints when comparing the fiducial setting (blue curve) and the Euclid / CSST resolution setting (orange / purple curves), but the improvement is not significant. Although a worse constraint is obtained with lower image resolution, we see that even with lower angular resolution we can distinguish between models with
$M_{\rm hf}$ of $10^7$ M$_\odot$ and $10^8$ M$_\odot$. The advantage
of Euclid and CSST will be sample sizes that are vastly larger than
50. Note that future Euclid/CSST observations will likely be different from the sample we simulate here: they will have, for example, different lens/source redshift distributions, data quality and Einstein radii. Those differences can potentially make our predictions here different from future real observations.

Similarly, we use average constraints above to demonstrate the result dependency on lensing configurations, while for the constraints from individual sets of 50 strong lensing systems with different lensing configurations, we refer the 2$\sigma$ limit histograms in Fig.~\ref{fig:hist_2sigma_multi}. Colors of the histograms follow the meaning of Fig.~\ref{fig:dependency}. The vertical dashed line mark the true input and the left regions below our test range are plotted in shadow. For each setting, we list the medians of 2$\sigma$ constraints in corresponding colors. As shown, the dependency reflected from the histograms is the same as that from the average constraints in Fig.~\ref{fig:dependency}. Among all the settings, the setting with a higher redshift source gives the best constraint, where when true $m_{\rm DM}=8.86$~keV, the median 2$\sigma$ constraint one could get is $m_{\rm DM} > 5.20$~keV.

To summarize, the tests shown here agree with the expectation that
lensing systems with longer exposure times, higher resolution, and
more low-mass dark matter haloes (higher source redshifts and a larger
area around the Einstein arcs) give tighter constraints.

\subsection{Model assumptions and limitations}
Our method makes a number of simplifying
assumptions which we now summarize and describe in more detail.  

By simulating low-mass dark matter haloes with the mass function of
Eq.~\ref{equ:mass_function} we neglect lensing perturbations from
subhaloes within the lens galaxy.  This is a good approximation since,
as \cite{Li2017} have shown, for realistic lensing configurations the
signal is dominated by line-of-sight haloes, rather than by subhaloes in
the lens. In any case, this assumption makes our results conservative,
as including subhaloes would boost the lensing perturbations,
increasing the signal we can extract from the lens model residuals. On
the other hand, there are uncertainties regarding subhaloes that do not
affect line-of-sight haloes, for example, the extent of tidal
disruption \citep{Despali2018,Richings2020}; marginalising over these
uncertainties would weaken the constraint on $M_\mathrm{hf}$.

We neglect any uncertainties on the amplitude and shape of the (CDM)
halo mass function. The amplitude is fixed by the value of the
cosmological parameter, $\sigma_8$, which is known to better than 1\%
from cosmic microwave background data \citep{Planck2020}. The shape is
known very precisely from cosmological simulations
\citep{Wang2020}. There is, however, some degeneracy between the
amplitude of the halo mass function and the half-mode mass. Some
lensing studies, such as that of \cite{Gilman2019} based on flux ratio
anomalies, have included the amplitude of the mass function as a free
parameter to be fit at the same time as the subhalo mass function.

In this work we have assumed an SIE lens mass model, whereas in
studies of real lenses a power-law mass model with external shear is
widely used \citep[e.g.][]{Vegetti2014, Dye2014}. This gives the mass
model more freedom to fit the residuals, reducing the signal left over
from substructures. Even then, the mass model may only provide an
approximate fit to the lens and result in residuals not associated
with subhaloes, at which point even more complex mass profiles
\citep{Nightingale2019}, or a potential corrections-based approach may
be required \citep{Vegetti2009}. We assume a core-Sersic model for the
source light, whereas studies using real data often use a
non-parametric approach that reconstructs a source's irregular
morphology \citep[see][]{Warren2003, Suyu2006}. \citet{Gilman2020d}
show how such an approach may absorb part of the residual signal of
the low-mass perturbers, reducing the available information on the
halo mass function. \text{PyAutoLens} has all the necessary
functionality to test these assumptions and this will be the topic of
future work.

Our method implicitly attributes all image residuals to the presence
of perturbing dark matter haloes when, in reality, there could be
other sources of mismatch between a true lens and source model and the
best-fit macro model. In the idealized setup used here we
have shown that this allows us to obtain a correct measurement of the
input half-mode mass. In reality, other sources of perturbations will
contribute to the power spectrum of the image residuals, including a
deficient model for the smooth component of the lens, an inaccurate
description of the telescope PSF and artefacts or correlated noises
introduced by the data reduction process. Not taking these effects
into account may bias our inference of the subhalo mass
function. Currently, the impact of any individual effect is unclear
and further investigation is required, noting that all effects that
impact the signal can, in principle, be included in our forward
modeling procedure and marginalized over.

On a positive note, our use of the power spectrum to define a summary
statistic to extract the signal from the residuals is likely to be 
suboptimal.  A more carefully crafted summary statistic or a machine
learning-based approach \citep[see][]{Brehmer2019, Rivero2020} can
potentially improve the signal that can be extracted from a lens
system and thus provide better constraints on $M_{\rm hf}$ than shown
here. It could also make the estimation less sensitive to systematic
sources of residuals as shown by \citet{Birrer2017}, and so increase
the constraining power of the forward modelling
method. 

\subsection{Comparison with flux ratio anomalies}

A theoretical investigation of the constraining power of flux ratio
anomalies was performed by \citet{Gilman2019}, who subsequently
applied their method to real observations, placing constraints on both
the dark matter mass function and the mass-concentration relation
\citep{Gilman2020a, Gilman2020b}. It is interesting to compare our
approach with that of \citet{Gilman2019}. Note, however, that, as we
just discussed, these authors treat the amplitude and slope of the
halo mass function as free parameters, whereas our tests assume a halo
mass function with only one free parameter, $M_{\rm hf}$. In our tests, for a universe with true
$M_{\rm hf}=10^7$ M$_\odot$, our fiducial calculation gives an upper
2$\sigma$ constraint (average constraint shown in Fig.~\ref{fig:accuracy}), $M_{\rm hf}= 10^{8.25}$~M$_\odot$, which is
slightly smaller than the upper constraints quoted by Gilman et al.,
$M_{\rm hf}= 10^{8.34}$~M$_\odot$, for a 2\% uncertainty on flux
measurements \citep[see Figure~8 of][]{Gilman2019}. Given the
different assumptions in the two studies this comparison serves to
show that the constraints that they provide are roughly comparable.

When considering the relative performance of our method and those based
on flux ratio anomalies it is pertinent to consider the mass scales
to which each method is sensitive. The ``coldest'' case in our tests
assumes $M_{\rm hf}=10^7$ M$_\odot$ and our signal is heavily
influenced by the lensing effects of larger perturbers with
$M_{200} \sim 10^{7-8}$ M$_\odot$. The small detection area of flux
ratio anomalies results in a lack of sensitivity to subhaloes with these relatively high masses (as these subhaloes are rare), but the ``point-like'' sources (with highly variable surface brightnesses over a small region) provide sensitivity
down to dark matter haloes with $M_{200} \sim 10^{6}$ M$_\odot$, a
scale that is not accessible with our approach. However, another problem faced by flux ratio anomalies is that the sparsity of information means that the degeneracy between anomalies caused by substructure and by a complex smooth halo is not easily broken, which can bias the inference on the abundance of substructures \citep{Xu2015, Hsueh2018}. The two approaches are
therefore complementary and, together, can provide constraints on the
halo mass function over a broad range of masses.


\section{Conclusions}\label{sec:conclusion}

The existence of a large population of dark matter haloes of all masses
down to subsolar values is a fundamental prediction of the CDM model
that distinguishes it from other currently popular models such as WDM
in which the halo mass function is truncated below a mass
$\sim 10^8$~M$_\odot$. Detecting the predicted population is a key
test of the standard cosmological model which would be ruled out if it
could conclusively be shown that low-mass haloes are absent or that
they are present in lower numbers than predicted.

In this work we have presented a method based on forward modelling
applied to the analysis of resolved strong lensing arc data that can
constrain a cutoff in the dark matter halo mass function. The key idea
is to simulate a large number of strong lensing images (forward
models) based on the best-fit macro model, with the addition of
perturbations to the lens model from low-mass dark matter haloes
(whose number depends on the assumed nature of the dark matter). These
images can then be fit in the same manner as the original observed
image, and the best-fit image residuals for the observed system can be
compared with the best-fit image residuals for the forward
models. This comparison is made by means of the power spectrum of the
image residuals within the Approximate Bayesian Computation
framework and leads to a posterior distribution for the half-mode
mass ($M_{\rm hf}$) which describes the cutoff in the halo mass
function. Our main results may be summarized as follows:

\begin{itemize}
\item For mock observed lenses constructed from parametric source
  light profiles and lens mass models, we confirm that information on
  low-mass dark matter haloes can be extracted from the power spectrum
  of the best-fit image residuals. As demonstrated in
  Fig.~\ref{fig:accuracy}, with a large number of observed lenses, a
  forward modelling procedure can correctly recover the true input
  half-mode mass, $M_{\rm hf}$ in average (the red curves). However, the scatter
  in the constraint is significant even though results for 50 systems
  have been combined together (blue curves), which may be attributed
  to the large intrinsic scatter in realisations of low-mass dark
  matter haloes. Taking the scatter into accounts, we compute the histograms of 2$\sigma$ limits (see Fig.~\ref{fig:hist_2sigma}) and find that, with 50 lensing systems,
  for our fiducial settings (see Table~\ref{tab:drawn_rges}), if the true (thermal) $m_{\rm DM}$ is 8.86 keV ($M_{\rm hf}$ = $10^7$
  M$_\odot$), the median 2$\sigma$ constraint on $m_{\rm DM}$ is that $m_{\rm DM} > 4.10$~keV (or there is 50\% chance that the $m_{\rm DM}$ can be constrained beterr than 4.10~keV at 2$\sigma$ level). Conversely, in a WDM universe 
  where the true (thermal) $m_{\rm DM}$ is 2.22 keV ($M_{\rm hf}$ = $10^9$
  M$_\odot$), one could get a median measure of $m_{\rm DM}$ to be between 1.43~keV and 3.21~keV at 2$\sigma$ level.

\item We have tested the dependency of the method on different lensing
  configurations and image quality settings. As shown in
  Fig.~\ref{fig:dependency}~\&~\ref{fig:hist_2sigma_multi}, the dependency agrees with expectations:
  higher redshift sources and/or larger areas around the Einstein
  arcs, and better data quality (longer exposure time and/or higher
  resolution) all result in a tighter constraint. Among our tests, the
  one with sources placed at $z=2.5$ produced the strongest
  constraint: in a universe with true $m_{\rm DM} = 8.86$ keV
  ($M_{\rm hf}=10^7$ M$_{\odot}$), particles with mass less than 5.20
  keV can be ruled out at the 2$\sigma$ level (see medians of the histograms in Fig.~\ref{fig:hist_2sigma_multi}). Although a slightly
  worse constraint is obtained from images with the resolution of
  Euclid/CSST, 50 systems can still provide a constraint on
  $M_{\rm hf}$, and with Euclid/CSST data we will have many more than
  50 strong lenses. We defer a thorough test using mock strong lensing samples similar to those expected from Euclid/CSST to future work.

\end{itemize}

Throughout this study we have made several simplifying assumptions,
particularly in using parametric models for the lens mass
distributions and the source light profiles. The effects of removing
these assumptions will need to be investigated before we can safely
apply our technique to real observations. We have shown that, in
principle, information on low-mass dark matter haloes can be
statistically extracted from the power spectrum of the image
residuals. In the near future, we hope that using a more powerful
summary statistic together with more advanced lens modelling
techniques (such as pixelized sources) can improve the power of this
technique. With thousands of new strong lensing observations expected
from future space telescopes, there is every prospect of pinning down
the mass of the dark matter particles using this forward modelling
technique.

\section*{Software Citations}

This work uses the following software packages:

\begin{itemize}

\item
\href{https://github.com/astropy/astropy}{\text{Astropy}}
\citep{astropy1, astropy2}

\item
\href{https://bitbucket.org/bdiemer/colossus/src/master/}{\text{Colossus}}
\citep{colossus}

\item
\href{https://github.com/dfm/corner.py}{\text{corner.py}}
\citep{corner2016}

\item
\href{https://github.com/steven-murray/hmf}{\text{hmf}}
\citep{Murray2013}

\item
\href{https://github.com/matplotlib/matplotlib}{\text{matplotlib}}
\citep{matplotlib}

\item
\href{numba` https://github.com/numba/numba}{\text{numba}}
\citep{numba}

\item
\href{https://github.com/numpy/numpy}{\text{NumPy}}
\citep{numpy}

\item
\href{https://github.com/rhayes777/PyAutoFit}{\text{PyAutoFit}}
\citep{pyautofit}

\item
\href{https://github.com/Jammy2211/PyAutoLens}{\text{PyAutoLens}}
\citep{Nightingale2015, pyautolens}

\item
\href{https://github.com/JohannesBuchner/PyMultiNest}{\text{PyMulitNest}}
\citep{multinest, pymultinest}

\item
\href{https://pythonhosted.org/PyQt-Fit/}{\text{PyQt-Fit}}
\citep{pyqtfit}

\item
\href{https://github.com/AshKelly/pyquad}{\text{pyquad}}
\citep{pyquad}

\item
\href{https://www.python.org/}{\text{Python}}
\citep{python}

\item
\href{https://github.com/scikit-image/scikit-image}{\text{scikit-image}}
\citep{scikit-image}

\item
\href{https://github.com/scikit-learn/scikit-learn}{\text{scikit-learn}}
\citep{scikit-learn}

\item
\href{https://github.com/scipy/scipy}{\text{Scipy}}
\citep{scipy}

\end{itemize}

\section*{Acknowledgements}

We thank the anonymous referee for useful suggestions which led to several improvements of the original darft. We also thank Nicola C.\ Amorisco for helpful discussions and suggestions, and Simon Birrer for insightful comments. QH, CSF and SMC are supported by ERC Advanced Investigator grant, DMIDAS
[GA 786910] and also by the STFC Consolidated Grant for Astronomy at
Durham [grant numbers ST/F001166/1, ST/I00162X/1, ST/P000541/1].  
This project has received funding from the European Union's Horizon 2020 research and innovation programme under grant agreement No.\ 776247.
RM acknowledges the support of a Royal Society
University Research Fellowship. RL acknowledge  the support of National Nature Science Foundation of China (Nos 11773032,12022306), the science research grants from the China Manned Space Project (No CMS-CSST-2021-B01).

This work used the DiRAC Data Centric system at Durham University,
operated by the Institute for Computational Cosmology on behalf of the
STFC DiRAC HPC Facility (www.dirac.ac.uk). This equipment was funded
by BIS National E-infrastructure capital grant ST/K00042X/1, STFC
capital grants ST/H008519/1 and ST/K00087X/1, STFC DiRAC Operations
grant ST/K003267/1 and Durham University. DiRAC is part of the
National E-Infrastructure.

\section*{Data availability}
The data underlying this article will be shared on reasonable request to the corresponding author.




\bibliographystyle{mnras}
\bibliography{example} 




\appendix

\section{Determining Multiplane Resolution}\label{AppendixA}

We simulate the effect of line-of-sight haloes by dividing the
lightcone from the observer to the source galaxy into a number of
intervals and approximating low-mass dark matter haloes within a 
a given redshift interval as lying on a single plane at the central
redshift. Reducing the redshift interval, $\Delta z$, between two
neighbouring planes (increasing the number of multiplanes) will
increase the accuracy of the approximation to the line-of-sight
effects, but will also increase the computational cost. The smallest
allowable $\Delta z$ should be determined by the accuracy one would
like to achieve in the inference of $M_{\rm hf}$. In practice, one
needs to make sure the ``difference'' caused by approximating
line-of-sight effects into multiplanes is smaller than the
``difference'' caused by changing $M_{\rm hf}$ by the amount one would
like to distinguish.

\begin{figure}
	\includegraphics[width=1.0\columnwidth]{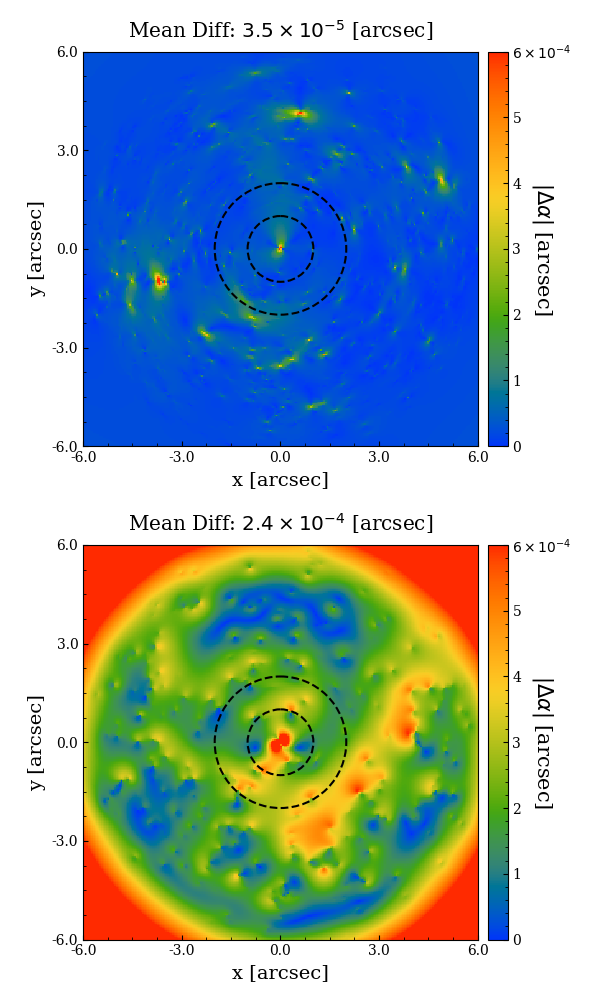}
        \caption{\textbf{Upper panel}: amplitude of the differences in
          deflection angles caused by changing $\Delta z$ from 0.01 to
          0.001. \textbf{Lower panel}: amplitude of the differences in
          deflection angles caused by changing $M_{\rm hf}$ from
          $10^7$ M$_\odot$ to $10^{7.1}$ M$_\odot$. The region between
          two dashed circles is the annulus between 1.0$^{\dprime}$
          and 2.0$^{\dprime}$. The unit of the colourbar is arcsecs.}
    \label{fig:diff_angles}
\end{figure}

Fig.~\ref{fig:diff_angles} shows an example of the amplitudes of the
differences in deflection angles when changing the values of
$\Delta z$ and $M_{\rm hf}$. The two panels are computed from the same
macro settings (a spherical isothermal sphere model with $R_{\rm E}$
of 1.5$^{\dprime}$ at $z$ = 0.5 and a source at $z$ = 1.0) and the
same low-mass dark matter haloes (generated from a distribution with
$M_{\rm hf}=10^7$ M$_\odot$). To simplify the test, instead of using
the mass function described in the main body of the paper, we impose a
``sharp cut'' on the mass function, such that no haloes are drawn with
$M_{200}$ smaller than $M_{\rm hf}$. Also, we assume that there is no
correlation between $M_{\rm hf}$ and the mass-concentration relation
to make sure massive haloes are unchanged when changing $M_{\rm hf}$.

The upper panel of Fig.~\ref{fig:diff_angles} shows the amplitudes of
the differences in deflection angles when changing $\Delta z$ from
0.01 to 0.001 (a very high resolution setting which we take as the
``accurate'' result). The lower panel shows the amplitudes of the
differences in deflection angles when changing $M_{\rm hf}$ from
$10^7$ M$_\odot$ to $10^{7.1}$ M$_\odot$. As seen from the figure, the
patterns of the differences caused by changing $\Delta z$ and
$M_{\rm hf}$ are different. We use the mean amplitude of the
difference in deflection angles within the annular region
1.0$^{\dprime}$ -- 2.0$^{\dprime}$ (the region between the two dashed
circles in the figure) to quantify the "differences". The mean
difference caused by changing $\Delta z$ is $3.47\times10^{-5}$
arcsec, which is smaller than the mean difference caused by changing
$M_{\rm hf}$ by 0.1 dex, $2.42\times10^{-4}$ arcsec, suggesting that
$\Delta z$ of 0.01 would not affect the ability to distinguish between
models with $M_{\rm hf} = 10^{7}$ M$_\odot$ and $10^{7.1} 
M_\odot$.

\begin{figure}
	\includegraphics[width=1.0\columnwidth]{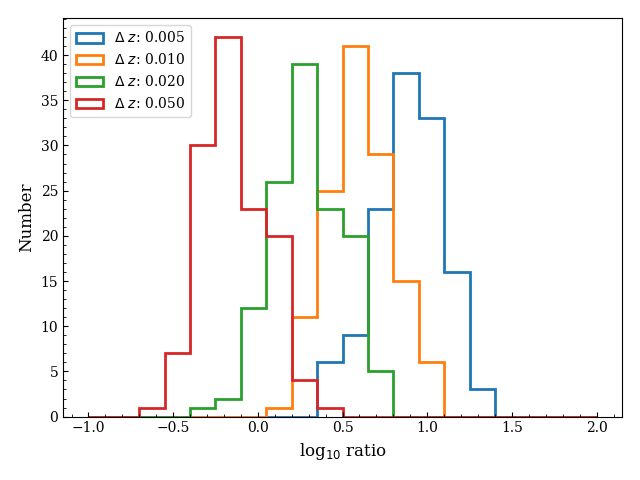}
        \caption{Histograms of the log$_{10}$ ratio of the mean
          amplitude of the differences in deflection angles caused by
          changing $\Delta z$ and $M_{\rm hf}$. Blue, orange, green
          and red colors shows results for $\Delta z$ = 0.005, 0.01,
          0.02 and 0.05 respectively.}
    \label{fig:diff_ratio}
\end{figure}

The results shown in Fig.~\ref{fig:diff_angles} are for a special
case, so we repeat the same procedure for 128 different realisations
of low-mass dark matter haloes. The orange histogram in
Fig.~\ref{fig:diff_ratio} shows the distribution of the log$_{10}$
ratio of the mean difference in amplitudes defined previously, that is
the ratio of the mean difference in the deflection angles between
models with $M_{\rm hf} = 10^{7}$ and M$_\odot = 10^{7.1}$ divided
by the mean difference in the deflection angles for the $\Delta z$
given in the legend. As shown by the orange histogram, the ratios from
all of the 128 realisations are larger than one, indicating that the
differences caused by approximating line-of-sight effects using
multiplanes separated by 0.01 in redshift is smaller than the
difference cause by changing $M_{\rm hf}$ by 0.1 dex. Also, we plot
the histogram of ratios for other values of $\Delta z$. As may be
seen, increasing $\Delta z$ increases the differences caused by
changing $\Delta z$ and, in some cases, the differences are larger
than those caused by changing $M_{\rm hf}$ by 0.1 dex. To summarize,
the value of $\Delta z$ we adopt for our tests, $\Delta z=0.01$, does
not affect our inference of $M_{\rm hf}$ to an accuracy of 0.1 dex.

\section{The distribution of constraints from sets of 50 lenses}\label{AppendixB}

To obtain an indication of the constraints on the DM particle mass that one expects from 50 lensing systems, in Fig.~\ref{fig:hist_2sigma} we plot histograms of the upper, or both upper and lower, 2$\sigma$ limits on $1/m_{\rm DM}$, for our fiducial lens and source properties. Each histogram shows the distribution of limits from 500 different realisations of a 50-lens sample, with upper limits shown when the true input, $M_{\rm hf}=10^{7, 8}$ M$_\odot$, while both upper (solid lines) and lower (dashed lines) limits are shown when the true input, $M_{\rm hf}=10^9$ M$_\odot$. 
For $M_{\rm hf}=10^8$~M$_\odot$, the probability that the true input will be above the 2$\sigma$ upper limit is $\sim 4$\%,  while for $M_{\rm hf}=10^9$~M$_\odot$, the true input is ruled out by the 2$\sigma$ limit $\sim$ 10\% of the time.\footnote{It may seem counterintuitive that the 2$\sigma$ limit rules out the true input with a probability different from $\sim 5$\%. However, here we are performing two different calculations: the 2$\sigma$ limits come from  Bayesian inference with a prior of $1/m_{\rm DM}$ assumed to be uniform over a certain range, while the multiple tests are only carried out for a particular value of  $1/m_{\rm DM}$. There is thus a mismatch between the prior for the inference and the tests. In other words, if one repeatedly computed 2$\sigma$ limits for a true $1/m_{\rm DM}$ uniformly distributed over the range, one would expect to find that the 2$\sigma$ limits rule out the truth with a probability of $\sim$ 5\%.} 
Additionally, in Fig.~\ref{fig:hist_2sigma_multi} we show histograms of upper, or both upper and lower, 2$\sigma$ limits from lens samples with different lensing configurations, corresponding to the different data quality and source and lens redshifts presented in Fig.~\ref{fig:dependency}.


\bsp	
\label{lastpage}
\end{document}